\documentclass[aps,prb,twocolumn]{revtex4-2}
\usepackage{graphicx}
\usepackage{epstopdf}
\usepackage{epsfig}
\usepackage{graphicx}
\usepackage{epsf,epic}
\usepackage{color}
\usepackage{subfig}
\usepackage{amsmath}
\usepackage{booktabs}
\usepackage{multirow}
\usepackage{physics}
\usepackage{hyperref}
\usepackage{amsfonts}
\usepackage{wrapfig}
\usepackage{breqn}
\usepackage{pstricks}
\usepackage[utf8x]{inputenc}
\usepackage{multirow}
\usepackage{fancyref}
\usepackage{pst-node}
\usepackage{float}
\usepackage{bm}
\usepackage{dcolumn}
\newcommand{\etal}{\textit{et al.\ }}
\newcommand{\ie}{\textit{i.e.\ }}

\makeatletter
\usepackage{etoolbox} % for \appto
\appto{\appendix}{%
	\@ifstar{\def\theequation@prefix{A.}}%
	{}%
}
\preto\maketitle{%
  \begingroup\lccode`~=`,
  \lowercase{\endgroup
  \let\saved@breqn@active@comma~% save breqn active comma
  \let~}\active@comma % set the active comma to what revtex4-1 wants
}
\appto\maketitle{%
  \begingroup\lccode`~=`,
  \lowercase{\endgroup
  \let~}\saved@breqn@active@comma % undo the change
}
\makeatother
\begin{document}
\title{Ultrathin 2D-oxides: a perspective on fabrication, structure, defect, transport, electron and phonon properties}
\author{Santosh Kumar Radha, Kyle Crowley, Brian A. Holler, Xuan P. A. Gao, Walter R. L. Lambrecht}
\affiliation{Department of Physics, Case Western Reserve University, 10900 Euclid Avenue, Cleveland, OH-44106-7079}
\author{Halyna Volkova, Marie-H\'el\`ene Berger}
\affiliation{MINES Paris, PSL University, Centre des Mat\'eriaux, CNRS UMR 7633, BP 87 91003 Evry Cedex, France}
\author{Emily Pentzer}
\affiliation{Texas A\&M University, College Station, TX 77843-3003}
\author{Kevin Pachuta, Alp Sehirlioglu}
\email{Corresponding author: alp.sehirlioglu@case.edu}
\affiliation{Department of Materials Science and Engineering, Case Western Reserve University, 10900 Euclid Avenue, Cleveland, OH-44106-7204}
\begin{abstract}
  In the field of atomically thin 2D materials, oxides are relatively
  unexplored in spite of the large number of layered oxide structures amenable to exfoliation. There is an increasing interest in ultra-thin film oxide
  nanostructures from applied points of view. In this perspective paper, recent progress in understanding the fundamental properties  of 2D oxides is
  discussed.  Two families of 2D oxides are considered: (1) 
  van der Waals bonded layered materials in which the transition metal is in
  its highest valence state (represented by V$_2$O$_5$ and MoO$_3$) and (2)
  layered materials with ionic bonding between positive alkali cation layers and negatively charged  transition metal oxide layers (LiCoO$_2$). The chemical exfoliation process and its combinaton with mechanical exfoliation are presented for  the latter.   Structural  phase stability of the resulting nanoflakes,
  the role of cation size and the importance of defects in oxides are
  discussed. Effects of two-dimensionality on phonons, electronic band
  structures and electronic screening  are  placed in the context of
  what is known on other 2D materials, such as transition metal
  dichalcogenides. Electronic structure is discussed at the level of many-body-perturbation theory using the quasiparticle self-consistent $GW$ method, 
  the accuracy of which is critically evaluated including effects of electron-hole interactions on screening and electron-phonon coupling. The predicted
  occurence of a two-dimensional electron gas on Li covered surfaces of LiCoO$_2$ and its relation to topological aspects of the band structure and bonding is presented as an example of the essential role of the surface
  in ultrathin materials. 
  Finally,
  some case studies of the electronic transport and the use of these
  oxides in nanoscale  field effect transistors are presented. 
\end{abstract}
\maketitle
\section{Introduction: Why 2D Oxides?} \label{sec:intro}
In the rapidly growing ``Flatland'' \cite{Abbott1884}
of 2D atomically thin materials, oxides
are still newcomers. After the Nobel prize was awarded to Novoselov and Geim
for  the discovery of
the remarkable properties of graphene,\cite{Novoselov04} several new
elemental 2D materials have been studied: black phosphoros,\cite{Li2014} antimonene and arsenene,\cite{Radhasb,Ares18review}
germanene
%\cite{Cahangirov09}
and silicene\cite{Sone2014}. Transition metal dichalcogenides (TMDC) rapidly gained
interests because of their remarkable valleytronic properties\cite{Mak10} and
the presence of a gap allowing field effect transistors\cite{Radisavljevic2011}
to be fabricated
to make use of the high-mobilities encountered in 2D materials. But in
spite of the large number of oxides that occur in layered crystal structures
and thus might be promising candidates for exfoliation, 2D oxides have
not yet been widely studied, especially from the piont of view of their fundamental properties
and prospects for electronic devices.
Oxides constitute  a broad  family of materials
with a wide range of potential applications, from catalysis to electronic,
photonic, ferroelectric, magnetic and multiferroic functionalities.
Understanding structure-property relations in free-standing, supported,
and confined two-dimensional ceramics or ``ceramic flatlands'' has been identified as one of the challenges for future work in ceramics by a recent National
Science Foundation workshop.\cite{nsfworkceram}

Nonetheless, over the last decade, there have been many reviews on 2D oxides
either as the main focus or as a part of a more general nano-materials
focused review papers or books. One of the earlier reviews of nanosheets
of oxides and hydroxides focused on synthesis, properties – especially photon-induced behavior - and their assembly, was published in 2010 by Ma \etal \cite{Ma10}, followed by a second review \cite{Ma15} as a part of a special issue on “2D Nanomaterials beyond Graphene” in 2015. Around the same time (2011)
another review by Mas-Balleste \etal\cite{Mas-Balleste11} also emphasized the importance and variety of 2D oxides.
A more recent focused review on 2D oxides can be found in 2019 paper of Hinterding \etal\cite{Hinterding19} introducing compositions, synthesis, microstructures and applications and in a 2018 paper of Uppuluri \etal\cite{Uppuluri18}
specifically related to the soft-chemistry of exfoliation.
In 2012, different morphologies of nanosheets were reviewed for Li ion storage by Liu \etal\cite{JLiu12}, summarizing nano-porous, sandwich-like, wave-like, flower-like nanosheets etc. Use of 2D-oxides as dielectric building blocks was reviewed by Osada \etal\cite{Osada12} in 2012. The review of liquid exfoliation of layered materials by Nicolosi \etal in 2013 while not focusing on oxides, included examples of exfoliation of oxides to form nanosheets.\cite{Nicolosi13}
A perspective on use of 2D nanosheets in hybrid systems, including oxides, was published in 2014 by Kim \etal\cite{Kim14} and was reviewed by Lee \etal in 2018\cite{Lee18}. Another review especially focused on variety of applications was published by Kalantar-zadeh\cite{Kalantarzadeh2016}. Other reviews and sections of books that introduce processing, and energy conversion and storage have been published over the years by ten Elshof \etal \cite{Elshof16}, Tan \etal\cite{Tan17}, Xiong \etal \cite{Xiong20}, Mahmood \etal \cite{Mahmood19} and Pang \etal \cite{Pang20}.
There has also been a lot of recent focus on the catalytic and environmental applications as reviewed by Haque \etal\cite{Haque17}, Heard \etal \cite{Heard19}, Zhang \etal \cite{Zhang20}, Safarpour \etal \cite{Safarpour20}.
Other applications of interest include sensors as reviewed by Shavanova \etal\cite{Shavanova16}, Lee  \etal \cite{Lee18}, Dral and ten Elshhof\cite{Dral18}. While not specifically on oxides, as a part of the review of 2D nanomaterials some information can also be found in the review papers by Tan \etal \cite{Tan17}, Jo \etal \cite{Jun19}, and Yang \etal \cite{FYang19}
Besides exfoliation, either mechanical or chemical, other routes to
formation of 2D ultrathin films or few-layer systems have been explored
using atomic layer deposition (ALD), and so on. 
An overview of 2D-oxdide synthesis methods and studies
was recently reviewed by Yang \etal.\cite{Yang19}.
Yang \etal \cite{yangJuan19} studied the formation of 2D oxide nanosheets
of CoO with nanoparticles as intermediate step. 
Xiao \etal reviewed layered cathode oxide materials for Na based 
battery materials. \cite{Xiao2020}
Barcaro \etal focused on ultrathin oxide layers grown on top of metals
and their structural motifs.\cite{Barcaro19}

From the above brief overview of the review literature it is clear that most papers had an application
oriented perspective. In the present perspective paper we take mostly a fundamental science point of view on 
the exfoliation process, the phase stability of 2D oxide nanosheets, their vibrational and electronic
properties that distinguish them from their bulk layered parent materials and how the above
affect their electrical transport. 

While a layered structure is a good starting point for fabricating
2D-oxides, the nature of the interlayer bonding is crucial for the success
of exfoliating. Two important classes of oxides can be distinguished.
Oxides like V$_2$O$_5$, MoO$_3$ have layered crystals structures in which
the layers are neutral and  the interlayer interactions are weak dispersive
van der Waals interactions. In these oxides, typically the transition metal
is in its highest valence state and perfectly balances  the oxygen ion charges. 
%Yet, several distinct crystal structures may exist even for the same
%compound chemical formula. 
%For example, V$_2$O$_5$ has $\alpha$,\cite{Bachmann61}, $\beta$\cite{Filonenko04} and $\gamma$ modifications
%\cite{Willinger04}
%where the $\beta$ form consists essentially of double layers in which vanadyl
%oxygens bonded to a single vanadium stick out on both sides of the double
%layer and the
%$\gamma$ and $\alpha$ phases differ by the way local motifs in the
%form of square pyramids (with one short V-vanadyl-O bond)
%are oriented relative to each other.
V$_2$O$_5$ has an additional interest in that 1D-chains occur inside the layers
and endow the material with properties in between 1D and 2D. As will be
detailed in Sec. \ref{sec:v2o5transport},
this leads to a high degree of  in-plane anisotropy in electronic transport\cite{Sucharitakul2017}.
Successful mechanical exfoliation using the ``scotch-tape'' method
has recently been accomplished and led to a start of the exploration
of this compound in 2D form. 

Orthorhombic $\alpha$-MoO$_3$ \cite{Seguin1995}
also has 
double layer character with oxygens bonded to a single Mo sticking
out from the double layer on either side and leading to weak interlayer
interactions. However, as the distance between layers is increased,
for example by inserting H$_2$O molecules,  its properties can be modified.\cite{Seguin1995}
A monoclinic form  is known in which the layers are just slided
slightly over each other laterally compared to the orthorhombic form
and this gives an indication that the properties
could be modulated by subtle changes in the interlayer interactons.
Ultrathin layers of MoO$_3$ have been successfully exfoliated and
high mobilities have been reported by some authors.\cite{Zhang2017}

The properties of these layered materials, which in their pure form
are fairly wide gap insulators, are strongly modified by defects, such as
oxygen vacancies or by intercalation with alkali or alkaline earth
atoms. This leads essentially to doping of the lowest conduction
bands of the insulator with electrons and turns them into n-doped
semiconductors. However, this also forms the basis for novel magnetic
properties. In particular in V$_2$O$_5$ intercalation  with Na in
the exact ratio of 1:1 forms the new compound NaV$_2$O$_5$ which is
antiferromagnetic.\cite{Carpy1972}
This is because $\alpha$-V$_2$O$_5$  has a 
narrow split-off conduction band separated from the higher conduction bands
by a small gap, which becomes half-filled when doped
with one Na per V$_2$O$_5$ unit.\cite{Bhandari15} This leads to the formation of
a magnetic moment, splitting the band in up and down spin, and subsequently
an antiferromagnetic coupling between moments along the direction of
the 1D chains. Besides this antiferromagnetic structure, another
phase transition, originally described as a spin-Peierls transition
occurs at low temperature but may in fact, be a combination of
charge disproportionation and spin-Peierls transition.\cite{Ueda96} This system
received a lot of attention in bulk form but ultrathin or monolayer
thick variants of it could offer a much finer control of the Na
or other intercalation content to further control and manipulate
these interesting magnetic phase transitions. The intercalated forms of
both V$_2$O$_5$ and MoO$_3$ are known as bronzes.\cite{Galy92,Stavenhagen1895} 
These materials have multiple applications in the field of catalysis,
are potential hosts for ions such as Li\cite{Whittingham04,DeJesus2016}
and Mg\cite{Tepavcevic15} in battery applications
and are also electrochromic.\cite{Panagopoulou19}
Their color and optical properties can be modified
by an applied voltage.

In this perspective paper, we will use both MoO$_3$ and V$_2$O$_5$
as representative examples of these van der Waals bonded oxides. We compare
their phonon properties and how they are influenced by the monolayering
in Sec. \ref{sec:phonons}. The electronic band structure of these oxides
is relatively well understood at the density functional theory level but
surprisingly, many-body-theory perturbation theory such as the $GW$ method
which should describe their electronic structure even more accurately
tends to overestimate the gaps. This is an impediment to  address the change
in gap in their monolayer forms because $GW$ self-energy effects are known
to have long-range effects in 2D materials.  We discuss our progress in this
respect in Sec. \ref{sec:bandgap}. 
Their transport properties are discussed in Sec. \ref{sec:transport}.

\begin{figure*}
  \includegraphics[width=\textwidth]{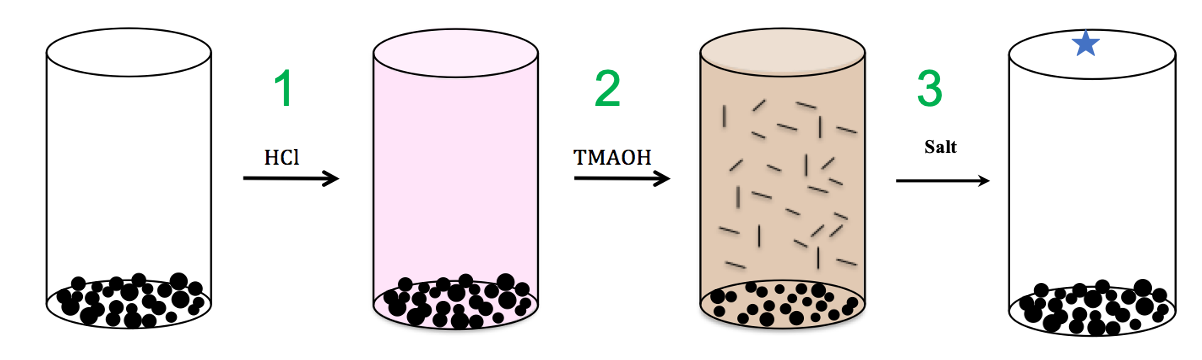}
  \includegraphics[width=\textwidth]{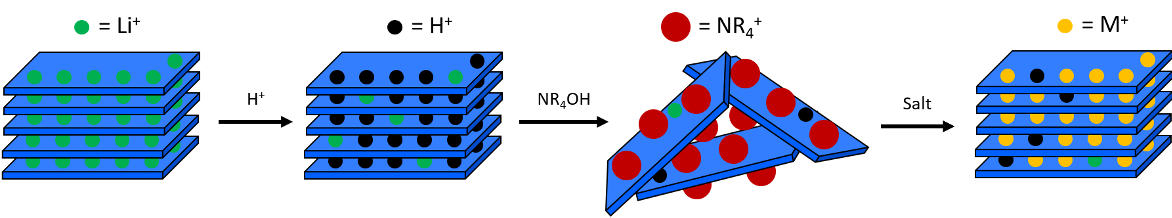}
  \caption{Steps of chemical exfoliation of CoO$_2$ nanosheets.\label{figexfo}}
\end{figure*}

On the other hand, many more oxides are essentially a layered arrangement
of different cations in a close packed oxygen lattice, one of them
a transition metal and the other a highly electropositive ion such as
alkaline ions, Li, Na, K. In these materials, the nature of the bonding
is such that the alkali atom donates its electron to the transition metal
oxide layer. We then have electrostatic bonding between negative transition
metal oxide layers and the positive Li ion layer. The prototypical
examples of this type of bonding are LiCoO$_2$ and NaCoO$_2$. The former
is the other main material discussed in this perspective paper. 
Exfoliation
of these materials is not as straightforward but has recently become
possible exploiting chemical ion replacement reactions. These are
described in some detail in Sec. \ref{sec:chemexfo}. The stability of the
resulting nanoflake materials under annealing is discussed in Sec. \ref{sec:anneal}.
A combination of chemical and mechanical exfolation provides a promising
route toward large area 2D mono or few layer thin forms of these
materials.\cite{Crowley2020,Volkova21} An important question that is not yet
fully answered is how much residual compensation
by alkali ions remains in operation and
how it modifies the LiCoxO$_2$ nanosheet properties. In LiCoO$_2$, the additional
electron supplied by the Li is precisely what is needed to give the Co
a Co$^{3+}$ formal valence state with a $d^6$ configuration which is
stabilized by  filling the lower energy $t_{2g}$  bands in
an octahedral environment leading to a low spin stable state.
This illustrates the subtle interplay between structure  and
electronic configuration. Removing part of the Li however,
now leads initially to a p-type semiconductor but with potential
Co$^{4+}$ localized ions if the Li-vacancies are ordered. The magnetic
moments of this $d^5$ configuration now come also into play.

The extreme case of CoO$_2$ layers is potentially a highly interesting
systems because the triangular lattice of Co spins is then highly frustrated
and has been speculated to be a candidate for the elusive spin-liquid state.
\cite{NingImai05,Yoshida20}
Remarkably, superconductivity has been claimed to occur in CoO$_2$ layers,\cite{Takada2003} not in isolated exfoliated form, but  when the layers are spaced by inserting a sufficient amount of water molecules inside the Na
layer of Na$_x$CoO$_2$. To what extent such a complex system
represents isolated 2D behavior of CoO$_2$ is not clear. 
However, while this provides a strong motivation to pursue exfoliation
of this system,  we are still a long way from perfect control over
this form of CoO$_2$. Numerous open questions start to emerge from
the exfoliation experiments carried out by our group. How much remaining
Li or other alkali ions remain attached to the layers as a result
of the chemical exfoliation and repricipitation route? How do they
influence the properties? How stable are the layers toward annealing
treatments, which may be required to make electrical contact to them?
Furthermore,  how are the very surfaces of CoO$_2$ covered with
Li different from the  bulk. Our initial findings on these  questions are discussed in later sections.
We discuss the effect of cation size on layered oxides of the LiCoO$_2$ type with different
cations replacing Li in Sec. \ref{sec:cationsize} as a way to increase the distance between the layers
and approach monolayer physics. We also include H in this study because of its role in the exfoliation
process. We return to the band structure of  LiCoO$_2$ in Sec. \ref{sec:bandgap} in the context of how
$GW$ theory applies to it and how we currently understand its optical properties. The surprising
prediction of a surface two-dimensional gas (2DEG) on LiCoO$_2$ Li terminated nanostructures
is discussed in Sec. \ref{sec:lco2deg} and its transport properties are presented in
Sec. \ref{sec:transport} along with those of MoO$_3$ and V$_2$O$_5$.

\section{Chemical Exfoliation of Li$_x$CoO$_2$ Nanosize Flakes}\label{sec:chemexfo}
The exfoliation of layered transition metal oxides (TMOs) has been relatively well studied, dating back to the 1990s and 2000s
beginning with the exfoliation of layered titanates and manganates.\cite{Sasaki98,Omomo2003} Since then, several layered TMOs
have been exfoliated using a soft-chemical exfoliation method.\cite{Ma10} It was not until 2009 that the exfoliation of
lithium cobalt oxide (LCO) was first reported (Kim \etal\cite{Kim09}). Since then only a few groups have reported further on this
topic\cite{Kim09,Edgehouse19,Pachuta19,Pachuta20,Li19,Adpakpang14,Crowley2020,Kim12,Jin15,Sasai17,Jin17,Volkova21,Jang19}
(many coming from our research group\cite{Edgehouse19,Pachuta19,Pachuta20,Crowley2020,Volkova21}
leaving the exfoliation of LCO relatively underexplored.

This soft-chemical exfoliation method can be described as a two-step acid-base reaction which swells and exfoliates layered TMOs to form
two-dimensional TMO nanosheets which can eventually also be reprecipitated. See Fig \ref{figexfo}.
First, the layered transition metal oxide is treated with a protic acid to replace the alkali metal
ions with protons in the interlayer. Second, this protic powder is subsequently reacted with a hydroxide bearing bulky molecule (e.g. tetramethylammonium hydroxide (TMAOH), tetrabutylammonium hydroxide (TBAOH)) to swell and exfoliate the layered structure into two-dimensional transition metal oxide nanosheets. While this method can be generalized for numerous layered TMOs, many factors that promote exfoliation are inherent to the properties of the layered TMO in combination with each subsequent chemical treatment that is utilized. In consequence, it is important to investigate and understand the material at each stage of the exfoliation process.

In the case of the exfoliation of LCO, reaction with protic acid solutions differ when compared to other more well-studied transition
metal oxides.\cite{Pachuta19} For instance, many layered TMOs can undergo complete alkali metal extraction and proton replacement while others, such as LCO, cannot.\cite{Kim09,Pachuta19,Pachuta20,Basch14}
In other words, alkali metal and proton concentrations in the powder vary depending on the protic acid treatment applied (e.g. acid composition, reaction duration, acid concentration, solution refreshment).
Moreover, when LCO is treated with high acid concentration solutions (e.g. 12 M hydrochloric acid),
the powder completely dissolves into ionic species. This instability is also reported at lower concentrations
(e.g. 0.1 M HCl – 3 M HCl) and plays a role in changing LCOs surface morphology, stoichiometry, cobalt and oxygen valence,
and defect concentration.
Due to the varying effect acid treatments have on LCO, exfoliation using a bulky hydroxide-bearing molecule (i.e. NR$_4$OH, where R = Methyl, Ethyl, Butyl) can have mixed results.\cite{Pachuta20}
For example, the concentration of NR$_4$OH needed to promote the highest degree of exfoliation has been demonstrated to be dependent on
acid treatment conditions. Furthermore, the degree to which exfoliation occurs is dependent on the ionic radius of the swelling molecule,
with TMAOH proving most effective. While several factors can affect the swelling and exfoliation behavior of LCO, the most important
of one is proton concentration in protic LCO. If this can be effectively quantified, the swelling and exfoliation behaviors of LCO
can be understood as well as other layered TMOs.

Along with the traditional soft-chemical exfoliation of LCO, other exfoliation and dispersion methods have been developed to more effectively utilize cobalt oxide nanosheets. These methods were created to avoid the destructive nature of the reagent required to promote exfoliation, TMAOH, which is highly caustic, a known etchant and developer of photoresist limiting the main applications of cobalt oxide nanosheets to electrophoretic deposition.\cite{Kim09} First, excess TMAOH can be removed from solution, through a solvent exchange after isolating cobalt oxide nanosheets from solution using high-speed centrifugation. The nanosheets can then be redispersed in water or another compatible solvent allowing for their use in non-aqueous applications. Second, cobalt oxide nanosheets can be precipitated/restacked from solution using salts. This opens the door, in effect, to synthesize oxide materials that are thermodynamically metastable (e.g. M$_x$CoO$_2$, where M is K, Ca, Al, etc. and found in the interstice of the layered structure). Finally, large and ultrathin cobalt oxide nanosheets in pH neutral solutions can be produced through washing the larger semi-exfoliated LCO particles (which are typically discarded as waste) initially isolated via low-speed centrifugation. These methods have allowed for a more versatile platform of the use of cobalt oxide nanosheets.

Although not statistically robust since the characterization is done using atomic force microscopy (AFM) and transmission electron microscopy (TEM), the cobalt oxide nanosheets produced using traditional exfoliation method have been demostrated to be a maximum of  
 0.4 $\mu$m$\times$0.4 $\mu$m in lateral size.\cite{Kim09,Adpakpang14,Jin17,Jang19} Moreover, there is little to no detail on the structure and properties of these cobalt oxide nanosheets making the process of modeling and understanding their fundamental properties a difficult task. New exfoliation methods have increased the size of cobalt oxide nanosheets and have allowed for advanced microscopic and electrical measurements to be performed advancing our knowledge of the structure and stability of these materials.\cite{Crowley2020,Volkova21}
\section{Structural Properties and Phase Changes Upon Annealing}\label{sec:anneal}
From the many years long interest in Li$_x$CoO$_2$ as a battery material,
it is well known that variations in Li content play an important role.\cite{Miyoshi18,Iwaya13,Motohashi09}
As the system is delithiated, at special simple fractional concentratons
of Li vacancies, such as $x=1/3$, $x=1/2$, $x=2/3$ ordering of the vacancies
may occur.\cite{Wolverton98}  
Also besides the layered $R\bar{3}m$ structure other forms of LiCoO$_2$
with a disordered or partially disordered arrangement of the Li and Co atoms
are known to occur at higher temperatures due to interdiffusion of the layers.
A  spinel type structure occurs with both Li and Co in octahedral sites
while the tetrahedral sites can stay empty \cite{Gummov93,Wang99}
and a fully disordered
rocksalt structure in which Li and Co randomly occupy octahedral sites has
also been reported. 

An important open question is to what extent the $R\bar{3}m$ structure
is stable in ultrathin nanosheets as function of temperature
and Li composition. A detailed electron microscopy study was recently
reported by our group for Li$_x$CoO$_2$ nanosheets  with $x\approx0.37$ obtained
by the chemical exfoliation route explained in Sec. \ref{sec:chemexfo}
and as function of annealing temperature.\cite{Volkova21}
The important findings are that as function of increasing  temperature,
a gradual disordering of the Li and Co atoms occurs.
Initially the structure of these nanosheets was found to consist
of a mixture of $R\bar{3}m$ and $C2/m$ regions where the latter
corresponds to a slight
distortion of the $R\bar{3}m$ structure related to the lower Li content.
Above 200$^\circ$C the $P2/m$ structure related diffraction spots
became evident and this indicates that around this temperature an ordering
of the Li vacancies is facilitated. At even higher temperature of
250$^\circ$C the spinel-type structure starts to appear, indicative of
disordering of Li and Co from their layered arrangement. Eventually above
350$^\circ$C, a fully disordered rocksalt structure was found.

\begin{figure}
  \includegraphics[width=5cm]{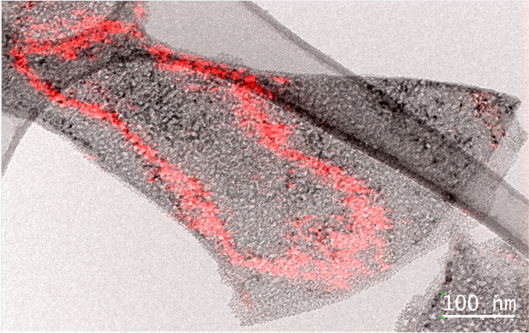}
  \caption{Superposition of bright (grey) and dark field image (red)
    of a strong  220 diffraction peak characteristic of the rocksalt
    domains of a LiCoO$_2$ nanoflake after annearling to 350$^\circ$C showing
    that rocksalt domains occur primarily at the rim of the flake.\label{fig:Li-loss}}
\end{figure}

The accompanying changes in electronic  properties were investigated via
electron loss spectroscopy in various energy ranges, the O-$K$ edge,
the Co-$L_2/L_3$ edges and the low energy interband transition range
up to the plasmon energy range. These provide information on the
degree of O-Co hybridization, the Co effective valence and so on.
The results were interpreted with the help of first-principles simulations
of these spectra. The main conclusions of this study were that in the first
temperature increase, the O-$2p$-Co-$3d$ hybridization is increasing but
then drops precipitously as the more disordered phases come into play.
In fact, a  more detailed analysis of the O-$K$ spectra, indicated that
the different shape of the spectrum at the highest temperatures
could only be explained by the pure rocksalt CoO phase.  This indicates
a loss of Li from those regions. Subsequently, it was found that 
those regions were predominantly located at the edges of the
nanoflake samples. This is documented here in Fig. \ref{fig:Li-loss}. The bright
domains in the dark field
image taken from  the strong 220 reflection  locate the rocksalt domains,
which are primarily found at the edges of the flake. This suggests
a gradual transitions from spinel to disordered rocksalt by loss of Li
from the edges of the flakes.

From the Co $L_2/L_3$ ratio it was concluded that the Co valence
initially increased, indicative of formation of some Co$^{4+}$ particularly
near the Li-vacancies when these become ordered, but formation of Co$^{2+}$
at the higher temperatures. The low energy loss spectra were found
to be in good agreement with a direct calculation of the loss spectrum
in the form of the imaginary part of the inverse of  the dielectric function
for small scattering vector ${\bf q}$ and also reproduced well the
plasmon region. The lowest energy transitions near 2-3 eV correspond to
transitions from the occupied $t_{2g}$ to the empty $e_g$ bands of
Co-$d$ electrons in their octahedral environment. The higher transitions,
near 7 eV were found to come from transitions from deeper valence bands with
a more dominant O-$2p$ character to the same lowest conduction band
rather than transitions from the top of the valence band to higher
Co-$4s$ or Li-$2s$ like states, which, in fact, occur at significantly
higher energy. 

The changes in structure, in particular when forms of CoO with
signficiant amounts of Co$^{2+}$ are present  were found to also
be accompanied by a drop in conductivity. This seems reasonable
because first of all disorder leads to increased scattering but
secondly in those phases the low spin optimal filling of the Co-$d$
bands for an octahedral environment are in competition with
other electron configurations which carry  a net magnetic moment
and exhibit a strongly correlated electronic structure.  In Ref. \cite{Volkova21}
this disordered paramagnetic state of CoO was modeled by means of
a special quasirandom structure (SQS) model 
with random placement of up spin and down spin magnetic moments
following the recently proposed polymorphous band structure
approach \cite{Trimarchi18}
for such correlated systems and using the DFT+U approach.
In essence this replaces the dynamic time dependent
fluctuations  of the spins by spatial fluctuations. 
  
The important take-away message from this study is that at elevated
temperatures, important structural changes can occur in nanoflakes,
their structure does not remain uniform and significant deviations
from the low spin LiCoO$_2$ band structure can occur. 
%\subsection{Other electron microscopy work?}
\section{Importance of Defects} \label{defects}
The direct effect of having 2D nanosheets is the greater surface to volume ratio achieved, and since the surface can be considerably different from the bulk it affects the concentration of ions with different valence, saturation, and coordination number, as well as the structure to decrease the surface energy, all to increase
the stability.\cite{Lei14,Chen11}
The defect concentrations and the related formation energy can change significantly and a minor type of point defect can become a dominant one or defect complexes can take over an isolated defect.\cite{Guan13} In addition, within the 2D material the behavior can be heterogeneous with edges behaving differently than the surface\cite{Nakada96,Huang14,Chiesa07}
or due to processing steps or post-processing treatment as was introduced earlier for annealed CoO$_2$
nanosheets (Sec. \ref{sec:anneal}). Since the performance of many nanomaterials strongly depends on the defect structure (e.g., vacancies, polarons etc.), and since processing techniques strongly affect the type, quantity and the distribution of defects, their characterization and control are critical and also a challenge. 

\begin{figure*}
  \includegraphics[width=8cm]{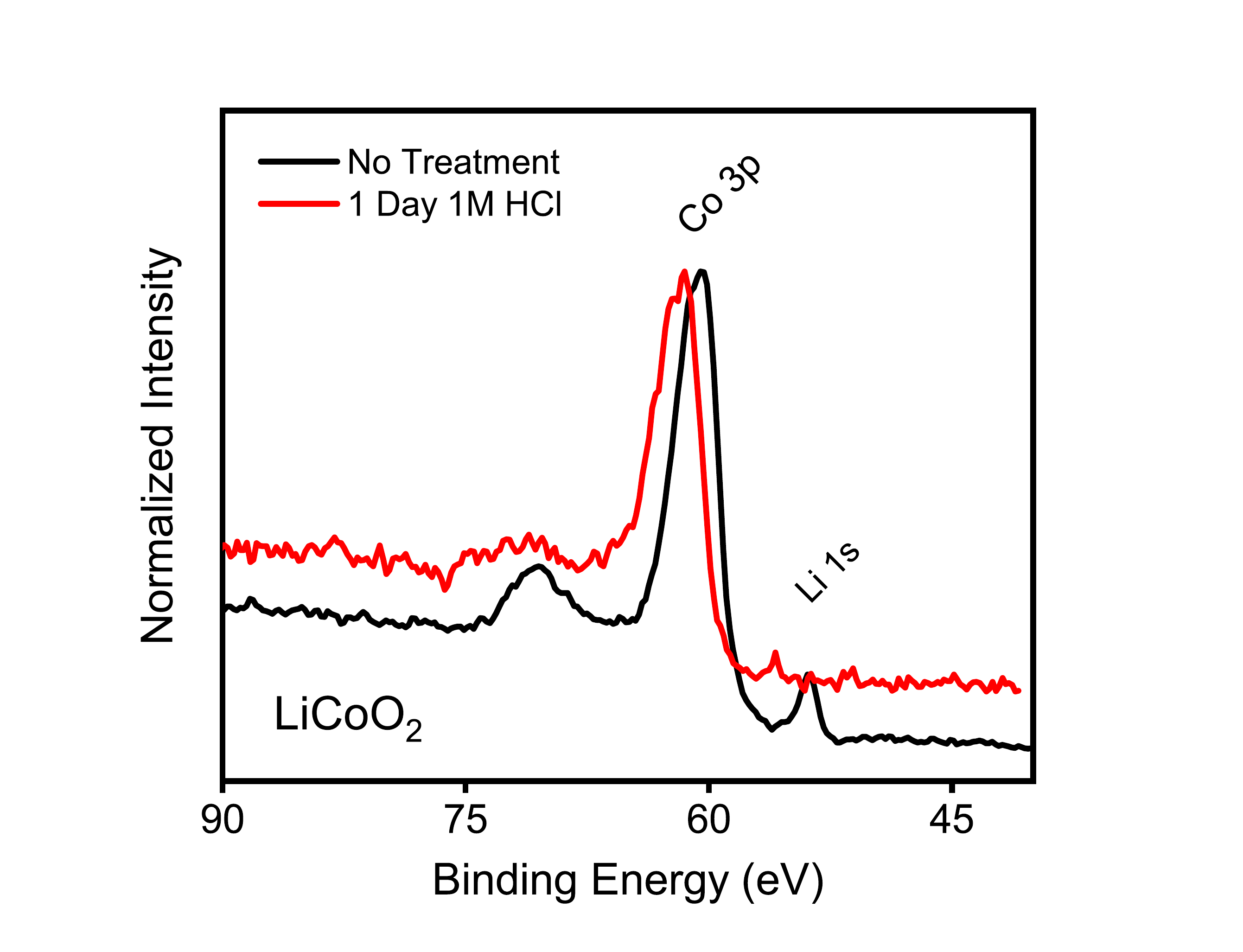}\includegraphics[width=8cm]{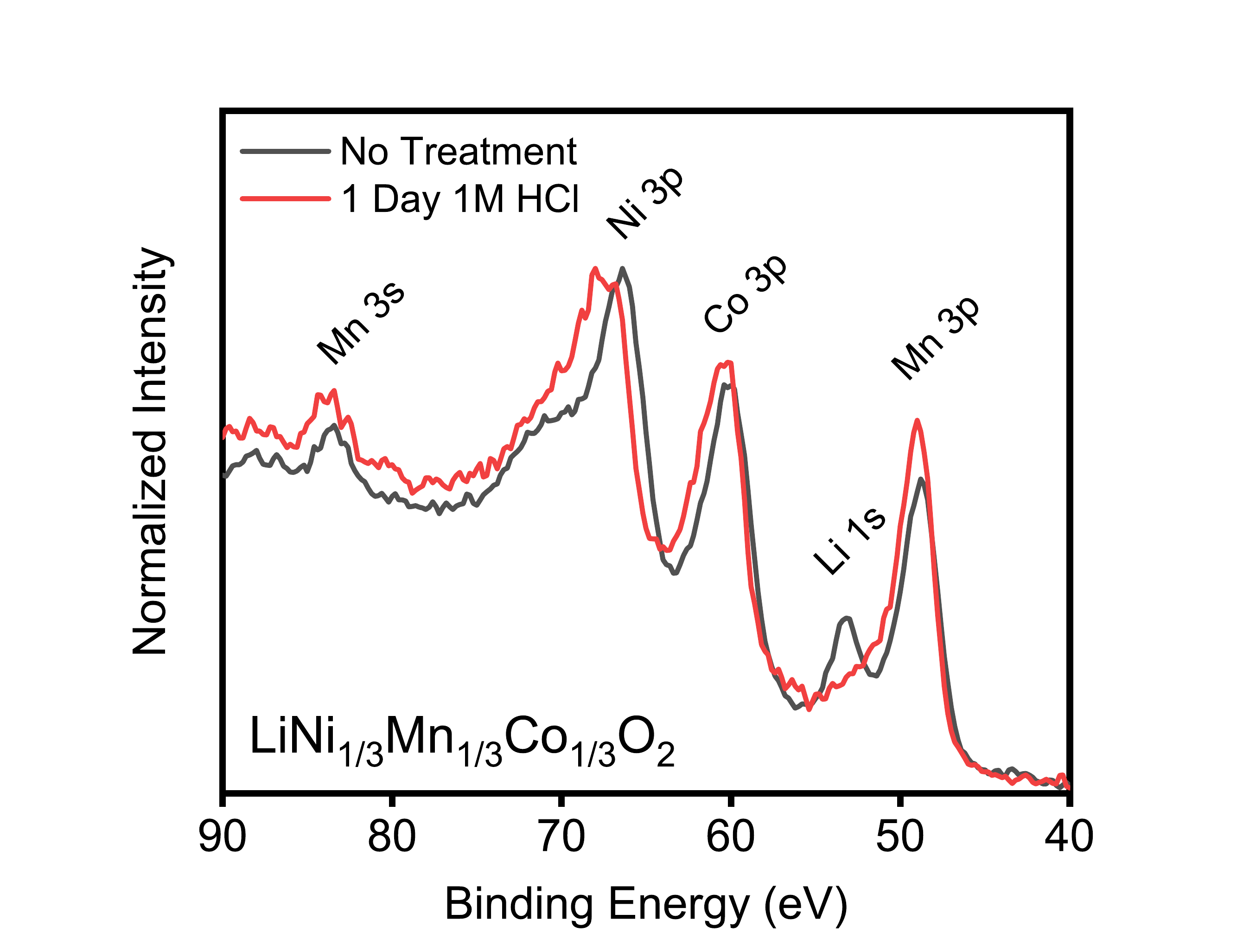}
  \caption{: (Left) XPS results for as-received LiCoO$_2$ (LCO) and 1M HCl treated LCO showing the removal of Li and the shift in Co-3p peak. (Right) XPS results for as-received Li(Ni,Mn,Co)O2 (NMC) and 1M HCl treated NMC showing the removal of Li and the shift in Ni-3p peak. Co and Mn peaks do not  change with treatment.\label{fig:nmc}}
\end{figure*}

The control of defect structures and defect structures themselves are a function of processing. Top-down approaches show promise in scaling-up for future production if a wide use of these nanosheets is targeted with inexpensive processing costs, and in obtaining nanosheets with large lateral sizes with lower structural defects needed for real devices without significant fabrication challenges. Although, top-down approaches limit the number of materials and systems that can be used (to layered structures), we are still far from realizing the full potential of them. There are also more recent techniques (i.e., adaptive ionic layer epitaxy - AILE)\cite{Wang2016} being developed to increase the lateral size of nanosheets during bottom-up processing which also opens up the material space to non-layered structures. 

Even in bulk, most oxides gain function due to their defect structures and chemically doped compositions with compensating electronic and ionic defects are used, instead of chemically pure and pristine compositions.\cite{Zunger03,Sehirlioglu10} Such defect engineering also applies to 2D oxides. When it comes to doping, in top-down approaches, compositions are most commonly designed at the bulk form that is exfoliated, however post-processing exposure to ions and solutions can also be used to exchange ions,\cite{Luo16,Okamoto11,Ida14}
while more flexibility exists for bottom-up approaches. A doped composition that retains the same layered structure can go through similar processing steps for isolation of 2D nanosheets. While such processing techniques can create similar nanosheets, doping elements can result in different defect structures. For example, Lithium cobalt oxide (LiCoO$_2$, LCO) and its relative, lithium nickel manganese cobalt oxide (LiNi$_{1/3}$Mn$_{1/3}$Co$_{1/3}$O$_2$, NMC), are both layered transition metal oxides (TMOs) utilized primarily as the active cathode for the Li-ion battery industry.\cite{Liuzhaolin99,Huddleston20}
NMC can also be exfoliated to form two-dimensional (2D) nanosheets using the same chemical treatments as described in Sec. \ref{sec:chemexfo}. However, how the material reacts to these chemical actions may vary. For instance, observable shifts in the binding energies of the transition metals of both transition metal oxides suggest a change in the oxidation state due to acid treatment as was discussed above in detail for LCO but differences lie in the binding energy shift behavior of each transition metal within LCO and NMC. LCO consists of only one transition metal (cobalt) therefore, cobalt shifts in binding energy. On the other hand, NMC consists of three transition metals (nickel, manganese, and cobalt) of which only nickel shifts in binding energy, Fig. \ref{fig:nmc}.
Both cobalt and manganese remain in a similar chemical bonding state before and after acid treatment, suggesting that nickel is the primary charge balancing ion associated with NMC after this chemical treatment. This also emphasizes that the change in the oxidation state of Co does not occur in the presence of Ni.

We can categorize the defects on the nanosheets into two: (i) defects that are similar to those of the surface defects in bulk materials such as changed coordination, buckling of the bonds, or presence of unsaturated atoms etc. and (ii) defects that form during processing due to the chemistry used such as preferential etching of cations. One can consider a third category for the surface defects as exfoliated nanosheets tend to have single termination while bulk termination can vary locally. The first category can be treated as the ``pristine'' nanosheet, and the changes in behavior compared to bulk can be sorted under the effect of increased surface to volume ratio. For example, in pristine SnO$_2$, nanosheet morphology accesses a larger number of four-coordinated surface Sn and two-coordinated surface O (rather than six and three, respectively, in the bulk) which form active sites for CO adsorption and subsequent oxidation.\cite{Sunyongfu13} Although it should be noted that, in this case, changes to the electronic structure in the confined structure (i.e, formation of increased density of states at the valence or conduction band edge), for example in comparison to nano-particles with the same surface area to volume ratio, results in additional factors that affect the efficiency of the reaction by providing enhanced charge carrier densities and fast ionic diffusion paths. In Co$_3$O$_4$, surface Co ions have lower coordination, in a similar fashion to other oxides such as MgO and TiO$_2$, thus the nanosheet form accesses a larger percentage of such Co ions.\cite{Chiesa07,Gaoshan16}
Lowered coordination can lead to more active sites in relationship to the environment. In the nanosheet form, higher density of states near the conduction band edge and higher charge concentration occurs in comparison to the bulk form. However, increasing the surface area does not always guarantee increased performance metric, as exemplified for catalytic activity of K$_4$Nb$_6$O$_{17}$.\cite{Sarahan08}
 
The second category is processing specific. Even when the same technique is used (i.e., protonation and exfoliation of LiCoO$_2$ with HCl and TMAOH, respectively), the defect chemistry can change with a change in a processing parameter such as molarity of HCl. The purpose of HCl treatment is exchange of Li$^+$ ions with H$^+$. However, the solution shows the presence of Co$^{2+}$ ions as characterized by UV-Vis. While some of that Co in the solution is due to dissolution of the formula units of LiCoO$_2$ as discussed in Sec.\ref{sec:chemexfo},  the quantification of the Li remaining in the powder and Co ions in the solution indicates leaching of Co ions into the solution beyond dissolution resulting in Co-vacancy defects in the structure to be exfoliated.\cite{Pachuta19} While the molarity and type of the acid used as well as the reaction time has an effect on the yield of exfoliation, they also vary the defect structure of the powders and thus can yield nanosheets with varying defect concentrations (here specifically Co-vacancies).\cite{Edgehouse19,Pachuta19}   

Many times, oxygen stoichiometry drives most performance characteristics and is generally achieved by post processing treatments of top-down processed oxide nanosheets; H$_2$ plasma treatment, UV radiation, and vacuum annealing.\cite{Geng18} In bottom-up approaches they can also be introduced during processing (e.g., sol-gel processing of TiO$_{2-d}$)\cite{Suriye08}. Even the difference between air and pure O$_2$, both of which are oxidizing atmospheres, can change the oxygen vacancy concentration both in bottom-up approaches\cite{Gao2017} and during phase transformation of existing nanosheets.\cite{Leifengcai14} As we have shown for CoO$_2$ nanosheets, however, as summarized in Sec. \ref{sec:anneal}, the changes in stoichiometry can be associated with changes in ordering as well as crystal structure. One also has to also consider that similar defects on the surface behave differently compared to bulk components. For example, in TiO$_2$, oxygen vacancies are compensated with two Ti$^{3+}$ cations which are symmetrically situated around the oxygen vacancies in bulk, while an asymmetric configuration occurs on the surface due to the presence of dangling bonds. This results in formation of one and two in-gap states, respectively.\cite{Liulianjun16}
Similar surface sensitivity due to vacancies exist in all materials; oxides\cite{Leeyueh-lin09},
selenides\cite{Zhaopeida14}, nitrides\cite{Abghoui16}, etc., however, in 2D materials the increased surface to volume ratio results in the dominant nature of the surface defects in the overall performance of the material. As noted before for SnO$_2$ with increasing density of states at the valence band edge in the 2D morphology, changes to the electronic structure, even if they do not have an effect in the mechanism, can have an effect on the efficiency and yield of the related reaction.\cite{Sunyongfu13}  

The importance of the oxygen vacancies in 2D oxide nanosheets can be exemplified by observations in materials systems such as TiO$_2$, ZnO, Co$_3$O$_4$, In$_2$O$_3$. Formation of disorder, and presence of oxygen vacancies and Ti$^{3+}$ cations on the surface in TiO$_2$ nanomaterials have been shown to increase visible and near-IR absorption, improved photocatalytic degradation of organic molecules, photoactivated reduction of CO$_2$, and H$_2$ evolution from water.\cite{Tanhuaqiao14,Zhuqing14,Sinhamahapatra15,Naldoni12,Chen11,Hanxiguang09}
Both the formation of defects and their related effects in performance are a function of crystallography, as formation energies of defects vary with the bonding environment and thus crystal planes. For example, for TiO$_2$, formation energy for oxygen vacancy is much lower in the $\{101\}$ planes. The anisotropy of the surface energies of crystallographic planes can also affect their
reactivity.\cite{Sunyongfu13} Reduction of TiO$_2$ nanoparticles to TiO$_{2-d}$, may result in formation of oxygen vacancies ($Vo^{..}$) coupled with Ti$^{3+}$/Ti$^{4+}$ with a ratio of $1:2$. However, it is also possible that
so-called Magneli phases form with formula Ti$_n$O$_{2n-1}$.\cite{Parras13}
%\textcolor{red}{MH thinks vacancies do not always form  Parass, J. Phys. Chem Lett. 13, 2171 (2013)} 
The presence of such vacancies results in in-gap states that increase absorption of visible light. Presence of surface oxygen vacancies can also act as trapping sites for electrons, effectively separating the photo-generated electrons and thus, limit recombination.\cite{Biwentuan14} Crystallography can also play an important role in separation of photoexcited electrons and holes, as they can transfer to different facets, critical in many applications related to catalysis.\cite{Sunyongfu13} In ZnO, oxygen vacancies, in a similar fashion, activated CO$_2$ reduction by providing an electron transfer. However, it was also shown that increasing oxygen vacancy concentration increased the activity of ZnO due to the enhanced binding strength of CO$_2$ to the surface and not due to the increase in the number of active sites.\cite{Geng18} In materials such as Co$_3$O$_4$, nanosheets can have oxygens with different coordination characteristics. Thus, the vacancy formation and its effects in performance can vary based on the type of oxygen site that becomes vacant.\cite{Gaoshan16} Oxygen vacancies, especially those near the surface, in
In$_2$O$_3$ nanosheets result in increased DOS both near CBM and VBM which both decreases the energy needed to excite electrons (due to DOS) as well as decrease the probability of recombination before the utilization of the photoelectron (due to position).\cite{Lei14}
 
The defects can be engineered for specific applications, such as providing active sites or increasing the binding energy for certain molecules.\cite{Hehaiying10} However, it can also lead to stabilization of adsorption of unwanted molecules that limit the targeted surface interactions.\cite{Green09} In addition, in the case of more than one type of oxygen site (based on its coordination), one type can preferentially be recovered by different molecules over the others.\cite{Schaub01}

\section{The Role of the Cation Size} \label{sec:cationsize}
An interesting question is whether the 2D properties of a metal oxide
layer can be already ascertained without exfoliation but simply by spacing
the layers further apart by increasing the size of the intercalation ions.
Although LiCoO$_2$ is not strictly an intercalated compound because the Li
constitutes an integral part of the bonding mechanism, we pursued this
idea computationally by investigating the properties of ACoO$_2$
with A  being Li, Na, K, Rb, Cs and H. H is included because the first step
in the exfoliation process is in fact, replacing Li by H.
At the same time, this study reveals some aspects of the role
of the size of the intercalating $A$-ion in the exfoliation process. 

\begin{figure}
  \includegraphics[width=8cm]{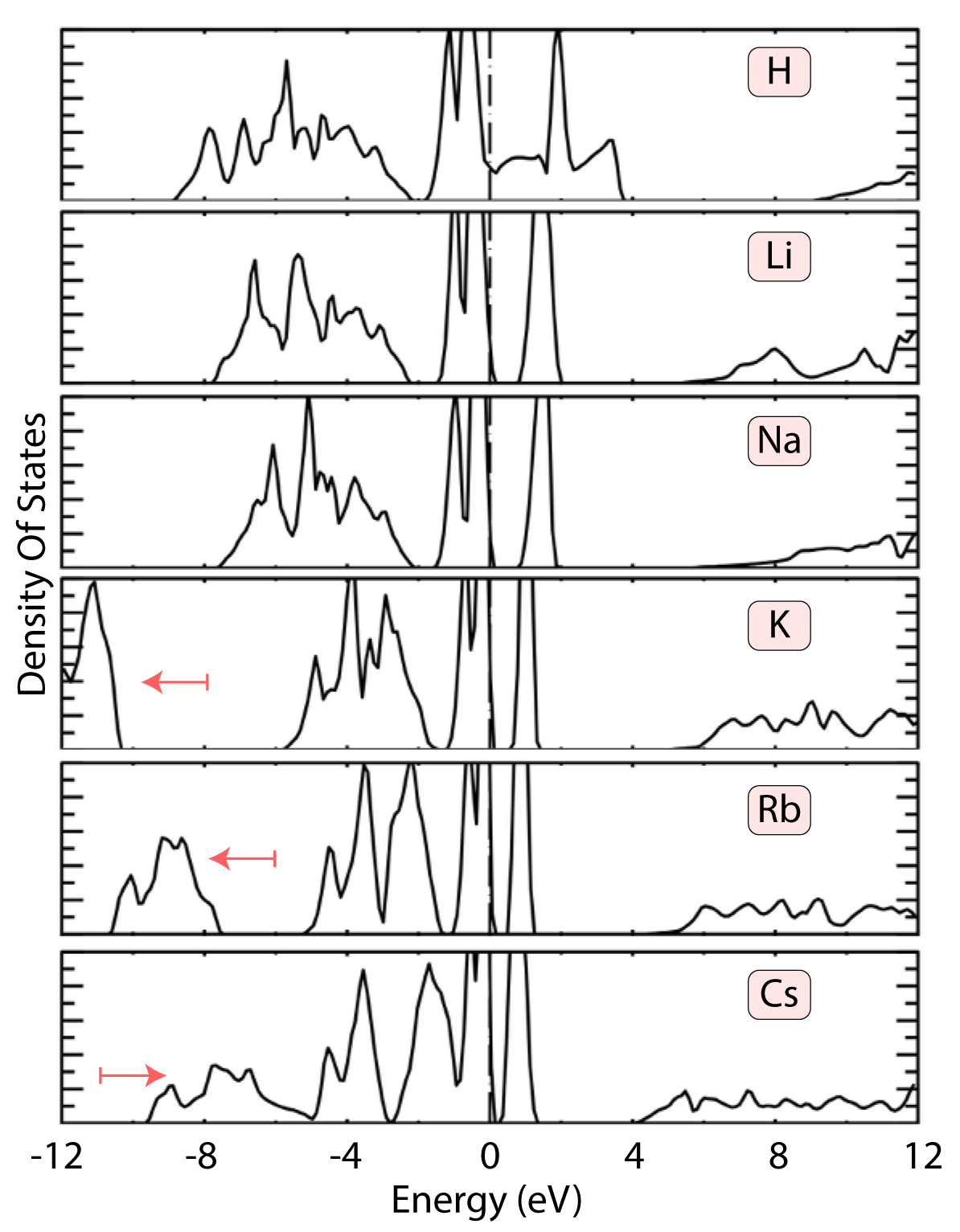}
  \caption{Density of states
    in $A$CoO$_2$  with  $A=H,Li,Na,K,Rb,Cs$
    in $R\bar{3}m$ structure 
    with in-plane lattice constant frozen at that of LiCoO$_2$
    and interplaner distances relaxed. The red arrows indicate the 
    semicore bands which are unphysically broadened and shifted up due to the
    unreasonable lateral proximity of the large $A$ atoms. \label{figdosion}}
\end{figure}
Our initial study started from the $R\bar{3}m$ structure of LiCoO$_2$
keeping  the lateral in-plane lattice constant fixed and simply replaced Li
by various other ions and relaxing the interplanar distance.
The densities of states resulting from this procedure are shown in
Fig. \ref{figdosion}.
We can see that not much change occurs when replacing Li by Na. Even replacing
Li by Na, Rb, Cs, the gap between the $t_{2g}$  valence band and
$e_g$ conduction band remains intact. However, one may notice a new
band at lower energy below the VBM and this then seems to push up the $t_{2g}$
and O-$2p$ derived bands. 
Further inspection of the partial
densities of states, reveals that these are related to the K-$3p$, Rb-$4p$ and
Cs-$5p$ semicore orbitals. Such a large band width for semicore orbitals is
unrealistic and is due to the fact that we did not allow the in-plane
lattice constant to relax in this calculation. However, if we allow the
in-plane lattice constant to expand, then the integrity of the CoO$_2$ layer
becomes questionable. 
This suggests, that there is limitation to this procedure to expand the
interlayer distance. In fact, a little bit of database search in
the Materials Project (MP) database,\cite{MP} shows that for K$_x$CoO$_2$ the
$R\bar{3}m$ like structure is only reported for $x=0.5$ but not for $x=1$.

The lowest energy structure of KCoO$_2$
has a $I\bar{4}$ space group with Co in tetrahedral coordination and the tetrahedra are connected in hexagonal rings. 
Another layered form with layers formed by
square pyramidal CoO$_2$ units is found in MP\cite{MP} with space group $P4/nmm$. 
This structure has a significantly
lower volume per formula unit  and is thus a phase that could
possibly be stabilized under high pressure. 

For Rb and Cs, the structures
are three dimensional with a  $Fd\bar{3}m$ space group. Both consist of
two interpenetrating diamond lattices of Rb (Cs) and CoO$_4$ tetrahedra.
The CoO$_4$ tetrahedral units are corner sharing. 
\begin{figure}
  \includegraphics[width=8cm]{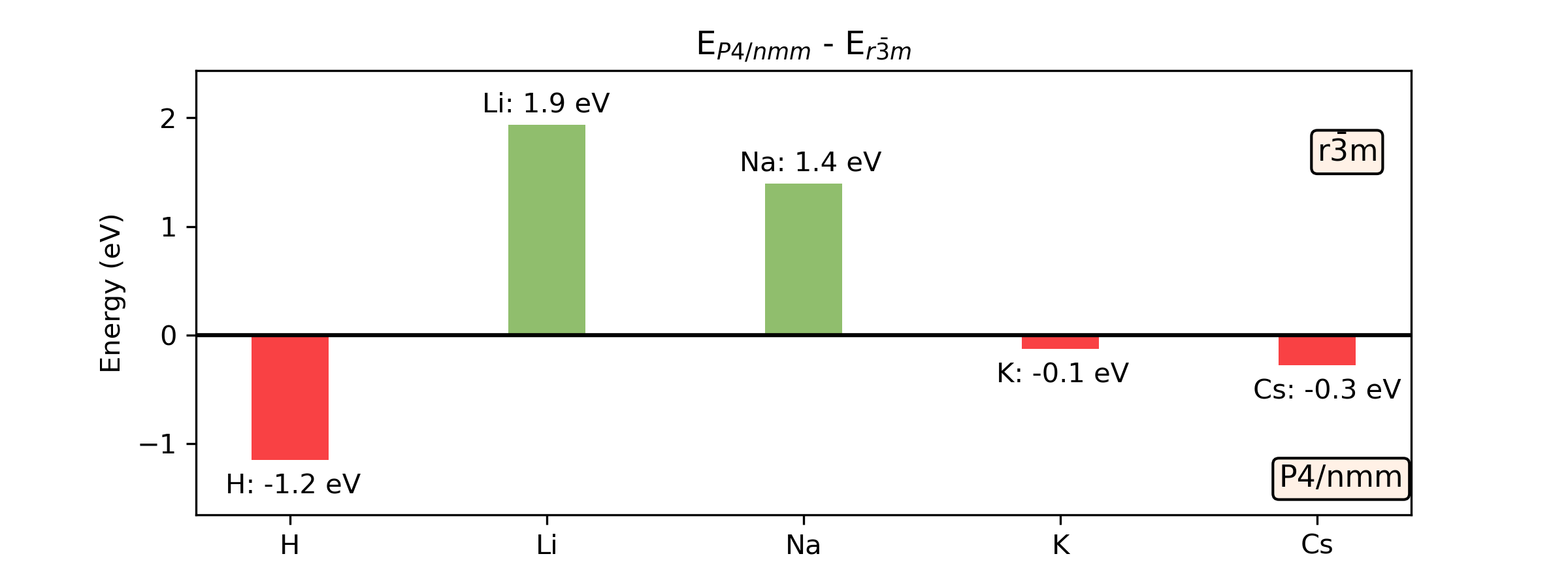}
  \caption{Total energy difference per formula unit between
    pyramidal coordination $P4/nmm$ and octahdedral coordination $R\bar{3}m$
    layered structures $A$CoO$_2$ for $A$ being H, Li, Na, K, Cs.\label{fig:totediffAcoO2}}
\end{figure}

Subsequently, we performed fully relaxed calculations of both the
$R\bar{3}m$ and $P4/nmm$   for LiCoO$_2$,  and replacing Li by H, Na, K
and Cs.  The results in Fig. \ref{fig:totediffAcoO2}
show that the $R\bar{3}m$ structure has lower
energy for Li and Na but the $P4/nmm$ structure has lower energy in the 
K and Cs case, and surprisingly also for H.
According to MP,\cite{MP} the $P4/nmm$ structure itself is 0.488 eV/atom higher
than the $I\bar{4}$ structure, which means 1.95 eV/formula unit.
Thus the $R\bar{3}m$ structure is unstable by more than 2 eV/formula unit
relative to its ground state $I\bar{4}$ structure for KCoO$_2$. 
The situation becomes worse for even larger alkali cations.
The result for H is somewhat surprising as the $R\bar{3}m$ structure
is listed as ground state structure  in MP.\cite{MP} It can however by
rationalized by the H forming an OH bond  with just  apical O of the CoO$_5$
pyramid instead of being shared equally between two CoO$_2$ layers.
Other forms with more pronounced OH bonds are found in MP.  As far as the
process of hydrogenation in the first step of the exfoliation procedure
is concerned, these calculations suggest that
the process is not as simple as just replacing Li by H because then
H would not have formed a simple OH bond but would be required to
stay halfway in between the layers.  However, Sec. \ref{sec:lco2deg}
shows that beyond a certain distance between the layers even Li
would undergo a symmetry breaking and associate with one CoO$_2$ layer.
We may expect the same for H if the distance between the layers is kept
higher perhaps because of
the aqueous environment. H$_2$O molecules inserted between the layer
may play a critical role in keeping the layers sufficiently apart
so that Li and H associate with one CoO$_2$ layer instead of two.
The critical role of the presence of H$_2$O is not yet fully
understood but is also found in the experimental studies as function
of solvent.\cite{Pachuta20} 

It is thus clear that the layered $R\bar{3}m$ structure is only
stable for relatively small alkali ions Li and Na. 
This may explain to some extent why inserting large organic TMA ions
breaks the structure apart and exfoliation occurs. It also
would appear that only a fraction of the Li can be replaced by TMA ions.

Finally what about the initial step of H replacement? In this case
we find a metallic band structure. Essentially, the H forms energy levels
within the $t_{2g}$-$e_g$ gap, which broaden into a band. This can
be explained by the lower energy position of the H-$1s$ state compared to the
Li-$2s$. In other words, H is not as electropositive.

%\textcolor{red}{Eventually some more information from Santosh
%  on gradually adding H and also on the bonding of H, how the O rotate
%  etc.}

\section{2D Effects on Phonons: V$_2$O$_5$ and MoO$_3$} \label{sec:phonons}
From earlier studies of 2D materials such as
TMDC it is clear that the reduction in dimensionality can have important effects on phonons. First of all the acoustic modes
show flexural modes which have a long wave dispersion proportional
to $q^2$ while in 3D the phonon frequencies behave $\omega\propto q$.
But optical modes in TMDC such as MoS$_2$
have also been found to show either red or blue shifts.\cite{MolinaSanchez11}
In this case it was found that the  out-of-plane, $A_{1g}$,  mode,
which becomes $A_1$ in the monolayer undergoes
a red shift, while the
in-plane $E_{2g}^1$ mode which becomes $E^\prime$ in the monolayer undergoes a
blue shift.  These shifts are of the order of a 2-4 cm$^{-1}$ or about 0.5 \%
of the mode frequency and have both
been observed experimentally and explained theoretically.  A redshift
might be expected for modes normal to the plane considering that removing
the interlayer interactions, however weak, should reduce the stiffness
of the system and thus decrease the mode frequency. However, for an in-plane
mode one would expect less of an effect of the interplanar force constants
because these would be shear forces. In any case, the blue shift seemed most puzzling but was explained by the effects of reduced screening in 2D on
the long-range dipolar force constants.

\begin{figure}
  \includegraphics[width=4cm]{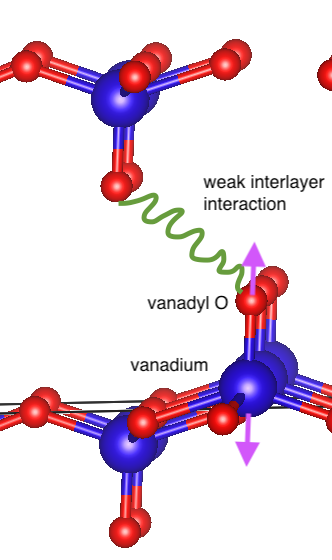}
  \caption{Vanadyl-oxygen vibration in V$_2$O$_5$ sketch.\label{fig:v-ovib}}
\end{figure}
\begin{figure}
  \includegraphics[width=8cm]{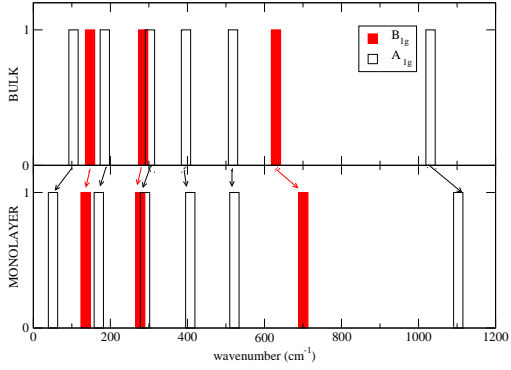}
  \caption{Phonon mode shifts betweenn bulk and monolayer V$_2$O$_5$.\label{fig:a1gb1gmono}}
\end{figure}

\begin{figure}
  \includegraphics[width=8cm]{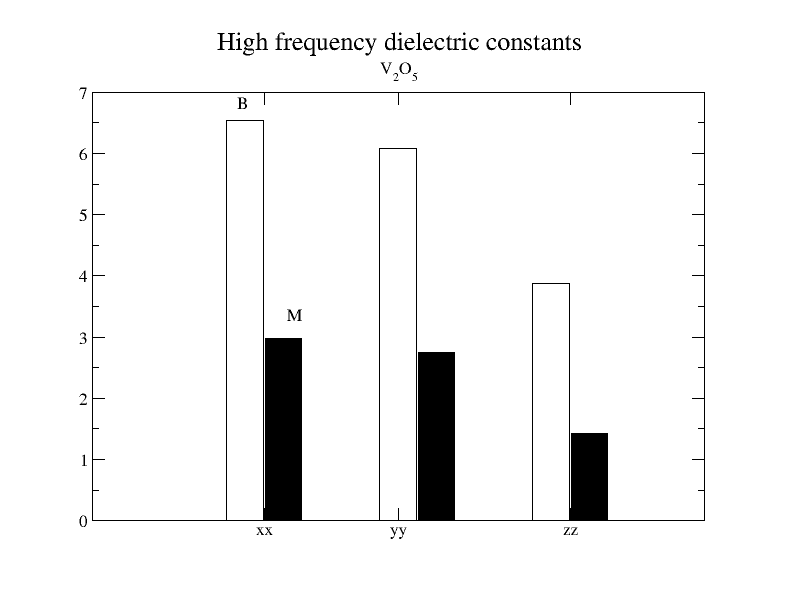}
  \caption{High frequency dielectric constants of V$_2$O$_5$ in bulk and
    monolayer.\label{fig:dielv2o5}}
\end{figure}

\begin{figure}
  \includegraphics[width=8cm]{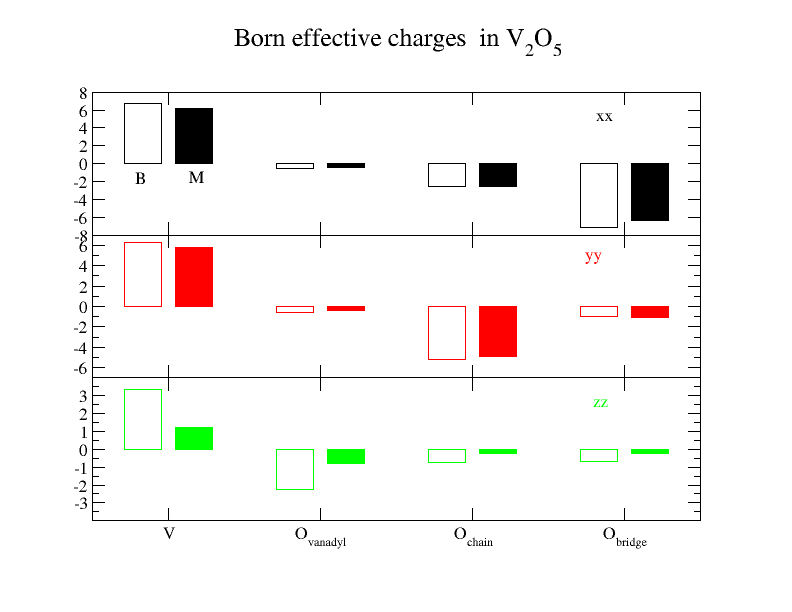}
  \caption{Born effective charges in bulk an monolayer V$_2$O$_5$, data from
    \cite{Bhandari14}, \label{fig:bornv2o5}}
\end{figure}

Similar effects have been found to occur for V$_2$O$_5$ \cite{Bhandari14}
but with a more interesting twist. In fact, the out-of plane modes
related to the bond stretch of the vanadyl-oxygen vanadium bond,
schematically shown in Fig. \ref{fig:v-ovib},
were found to undergo a blue shift in the monolayer compared to the bulk.
Fig. \ref{fig:a1gb1gmono}  shows the $a_{1g}$ and $b_{1g}$
Raman active mode shifts from bulk to monolayer in V$_2$O$_5$. 
Applying a similar analysis to that of Molina-S\'anchez and Wirtz
\cite{MolinaSanchez11} it was found that the applicable long-range force constant was in fact
decreased  instead of increased because  not only the dielectric constants
(shown in Fig.\ref{fig:dielv2o5})
were reduced in the monolayer but also the relevant Born effective charges
were reduced by a similar factor. See Fig. \ref{fig:bornv2o5}
for the Born charges, note that for the vanadyl oxygen only the $zz$ component
is large because it is bonded in the $z$ direction
while for the bridge oxygen only the $xx$ component is large
because they are bonded only in the $x$ and for the chain oxygen, both
$xx$ and $yy$ are sizable because these oxygens are bonded in $x$ and $y$
directions. However, one can also see that the V $xx$ and $yy$ components
are not much changed while the $zz$ component is reduced by almost
a factor 3. 
Since the dipolar force constants are proportional to the
product of the Born charges of the two atoms involved in the
bond stretch and inversely proportional to  the in-plane dielectric constant,
the long-range force constant is reduced.  So, why was a blue shift obtained?
The other  difference with the MoS$_2$ case is that here a V-O bond stretch
is at play rather than a Mo-Mo bond stretch in the plane. Thus the
Born charges have opposite sign and this results in the long-range force
constant having opposite sign to the short-range force constant. The latter
barely changes and so the net result of a decreased long-range force constant
is less opposition to the short-range force constant. The latter then becomes
dominant in the monolayer and explained the blue shift.
Some other modes in V$_2$O$_5$ were also found to have red-shifts
and although a detailed analytical insight is only possible for
some of the simpler mode patterns, these effects still remain as predictions.
Although thin V$_2$O$_5$ layers (down to about $\sim 10$ nm or 20 layers)
have been exfoliated,\cite{Sucharitakul2017}  the monolayer limit
has not yet been achieved, so that these predictions remain to be confirmed
by experiment.  Very recently,  a chemical exfolation study of V$_2$O$_5$
by Reshma \etal\cite{Reshma21} reported bilayer V$_2$O$_5$. It confirms
the blue shift of several vibrational modes, in particular the high-frequency
$A_{1g}$ mode, in qualitative agreement with the theoretical predictions.
It also predicted a significant increas in band gap from 2.39 eV in bulk
to 3.72 eV in the bilayer to be discussed in the next section. 
 
More recently, we have also studied such shifts on vibrational modes
in MoO$_3$.\cite{Amol2020} Again, the story turns out to be a bit different
than in V$_2$O$_5$. The most interesting modes to analyze again involve
a bond stretch of the short Mo-O$^{(2)}$ bond, which occur at
about 1000 cm$^{-1}$. The same analysis  of the dipolar force-constants
apply and lead to a net but small increase of the  force constant
directly between Mo and the O$^{(2)}$ bonded to it. However, in contrast
to V$_2$O$_5$, the geometry of the layers leads to a stronger interlayer
O$^{(2)}$-O$^{(2)\prime}$ interactions between these O in adjacent layers.
They just happen to be closer to each other laterally and hence there
remains a larger interlayer force in bulk. This effect turns out to dominate
the blue shift tendency arising from the Born charge and dielectric constant
reductions, and thus, ultimately lead to a red-shift.
In absolute value this shift is nonetheless larger than in MoS$_2$,
it is about 14 cm$^{-1}$ or 1.3 \% and should thus be easy to
observe by Raman spectroscopy.

In summary of this discussion, changes in screening and Born effective
charges which related to the polarizability of the bonds in 2D
were found in various 2D materials to lead to   interesting shifts
in phonon frequencies but the detailed application of these ideas
is rather subtle and requires a detailed analysis of the force constants.
It also shows that in the bulk of these layered materials, the
interlayer forces, even when weak still affect vibrational modes.

Another general effect of the dimensionality on infrared active phonons
was pointed out by Sohier \etal\cite{Sohier17}. In 2D the long-range limit 
of the in-plane dielectric function \cite{Cudazzo11}
$\varepsilon_{2D}({\bf q}_\parallel)=\varepsilon_m+r_{eff}|{\bf q}_\parallel|$
(with $r_{eff}$ the effective screening length)
becomes only the $\varepsilon_m$ of the surrounding medium. This is a clear
dimensionality effect: the field screening is dominated by the
electric field lines that go through
the medium outside instead of inside the 2D material.  On the other
hand at short distance (large $|{\bf q}_\parallel|$) the screening becomes
proportional to  $|{\bf q}_\parallel|$.  This then affects the LO-TO splitting
in a qualitative manner. 
First of all the Coulomb interacton in 2D is $2\pi/|{\bf q}_\parallel|$
instead of $4\pi/|{\bf q}|^2$ in 3D. Secondly the essential dependence of
the screening on wave vector implies that the LO-TO splitting will
go to zero for ${\bf q}_\parallel\rightarrow 0$. In practice, there
is a small region of order $1/r_{eff}$ where the LO-TO splitting
increases linearly with $|{\bf q}_\parallel|$.
These effects may manifest themselves more strongly in 2D oxides
where the ionic nature of the bonding leads to relatively low
electronic (high-frequency) dielectric screening to begin with. 

Apart from these specific 2D related changes in phonon modes,
vibrational spectra, specifically Raman spectroscopy has been found useful
as a characterization tool for 2D nanoflakes. The in-plane anisotropy
of the modes can be used to determine the crystal axes of a nanoflake
in cases where XRD would be difficult to carry out. This method was
used for example by Sucharitakul \etal \cite{Sucharitakul2017} to determine
that the long axis of the rectangular nanoflakes obtained
by mechanical exfoliation corresponded to the direction of the chains
in the structure.

\section{2D Effects on Electronic Structure} \label{sec:bandgap}
\subsection{Accuracy of $GW$-methods for layered oxides} 
Dimensional effects on electronic structure are well known to be important.
The basic model of quantum confinement goes back to the quantum mechanics
textbook example of a particle in a potential well. Discrete bound states
occur in the well  and the separation between the levels increases as the
the potential well becomes narrower, essentially $\propto1/w^2$ for an
infinite square well with width $w$. This already formed the basis
for the use of semiconductor heterostructures long before the
advent of 2D materials. On the other hand, for 2D few layer materials,
which are usually obtained from a weakly interacting layered material
to begin with, it is not so obvious why breaking those weak interactions
would strongly affect the electronic structure of the layer itself.
One would expect that the confinement effects are already present in
the layer even if we have a stack of such weakly interacting layers.

Yet, graphene is clearly different from graphite in the sense that
the $\pi$ electrons now form a Dirac cone with linear band dispersion and 
even bilayer graphene is essentially distinct from single layer graphene
in that a gap can open. For transition metal dichalcogenides such as
MoS$_2$ it was found early in the development of 2D materials, that
the nature of the gap can change from indirect to direct when going from
bulk to few layer or monolayer systems. This is because specific
states in {\bf k}-space have different deformation potential
and  react less or more strongly to the interlayer spacing depending
on their atomic orbital composition, which may point more in-plane
or out-of-plane.  To discover such effects, reliable prediction
of the electronic structure is a prerequisite.

It is well known that band gaps are typically underestimated by
local or semilocal exchange-correlation functionals in density
functional theory. The quasiparticle self-consistent $GW$ method (QS$GW$),\cite{MvSQSGWprl,Kotani07} on the other hand provides accurate results for most materials,
including all 
tetrahedral semiconductors. However, how well this approach works
for often more strongly correlated transition metal oxides is not yet
clear.  We here briefly discuss the essential concepts of the QS$GW$ method
before we discuss how it applies to the oxide materials of interestin in this paper.

In Hedin's \cite{Hedin65,Hedin69}
$GW$ approximation to the dynamical self-energy in many-body-perturbation
theory, $G$ stands for the one-electron Green's function and
the self-energy $\Sigma(1,2)=iG(1,2)W(1+,2)$ where $1\equiv\{{\bf r}_1,\sigma_1,t_1\}$ is a shorthand for position, spin and time of the particle labeled $1$.

In the  most prevalent implementation of the $GW$ method, the
$G^0W^0$ approximation, 
the quasiparticle excitation energy equation,
\begin{equation}
  [H^0-v_{xc}({\bf r})]\psi_n({\bf r})
  +\int d^3r^\prime \Sigma({\bf r},{\bf r}^\prime,E_n)\psi({\bf r}^\prime)=E_n\psi_n({\bf r}), 
\end{equation}
is  solved by first-order perturbation theory starting from
the independent particle Kohn-Sham Schr\"odinger equation
\begin{equation}
  H^0\phi_n({\bf r})=\epsilon_n\phi_n({\bf r}),
\end{equation}
by assuming $\psi_n\approx\phi_n$ and calculating the one particle
Green's funcion $G^0$ and polarizaton $P^0(1,2)=-iG^0(1,2)G^0(2,1)$
and hence screened
Coulomb energy $W^0(1,2)=v(1,2)+\int d(3,4) v(1,3)P^0(3,4)W^0(3,4)$,
or symbolically $W=[1-vP]^{-1}v=\varepsilon^{-1}v$, 
using the Kohn-Sham
$\phi_n$ and $\epsilon_n$. Essentially, the self-energy $\Sigma({\bf r},{\bf r}^\prime,E_n)$ replaces the exchange-correlation potential $v_{xc}({\bf r})$.
In self-consistent $GW$ theory,
one would need so solve the equations $G=G^0+G^0\Sigma G$,
$W=v+vPW$, along with the definitions of $\Sigma=iGW$ and $P=-iGG$
consistently. Here we have dropped the variable dependencies for simplicity, 
which however hides that these are in fact integral equations, which
in practice are solved by means of a basis set expansion and after
transforming to the Bloch function {\bf k}-space and energy domain. 

On the other hand, in
the QS$GW$ method, one recognizes that
perturbation theory works best if the perturbation is small, in other words
when the $\Sigma$ is in some sense close to $v_{xc}$. One thus 
adds  $\tilde\Sigma-v_{xc}^0$ to the $H^0$ Hamiltonian, with
the energy-independent, Hermitian 
$\tilde\Sigma_{ij}=\frac{1}{2}\mathrm{Re}\{\Sigma_{ij}(\epsilon_i)+\Sigma_{ij}(\epsilon_j)\}$ and iterates  this to self-consistency.
Here the $\Sigma$ is represented by a matrix in the basis of the $H^0$
eigenstates. The result is that the QS$GW$ method becomes
independent of the starting $H^0$ approximation, which is then usually taken to be the LDA or GGA. Typically QS$GW$ slightly overestimates the gap
and this has been largely attributed to the lack of electron-hole
interactions in the calculation of the screened Coulomb interaction $W$
in the standardly used random phase approximation (RPA).
The upshot is that with a small and systematic
correction for the underscreening, \ie
only including 80 \% of the $\tilde\Sigma$, the quasiparticle energies
agree typically with experiment within $\sim0.1$ eV.
This 80\% rule or $0.8\Sigma$ approximation
can be further justified\cite{Bhandari18}
or avoided by including the ladder diagrams,
which represent electron-hole interactions 
in the calculation of $W$ within the Bethe-Salpeter-equation (BSE) approach,\cite{Cunningham18} or via a detour to time-dependent density functional theory
(TDDFT) using an appropriate kernel \cite{Shishkin07,ChenPasquarello}
but this is still significantly more  computationally demanding. 

Having explained the methodology briefly, and referring the reader
to the literature quoted above for details, we now ask ourselves how well
does this approach work for the layered oxides we are interested in and
how do they work of few-layer 2D systems?
Surprisingly, in V$_2$O$_5$ we found that the  band gap
in the more advanced QS$GW$ method strongly overestimates the gap.
While the lowest direct gap is 2.0 eV in LDA, the corresponding
QS$GW$ gap is 4.45 eV (3.96 eV in $0.8\Sigma$), much larger
than the experimental value of 2.35 eV. The direct gap at $\Gamma$ is
2.30 eV and the lowest indirect gap is 1.73 eV.
Given that the experimental value extracted from optical absorption
most likely corresponds to the lowest direct gap, the LDA, while still underestimating, comes actually closer to the experiment than the QS$GW$ method.

Turning now  to the monolayer, we find that the LDA gap only moderately
increases in the monolayer compared to the bulk by about 0.5 eV
and this is mainly due to the reduced dispersion of the bands along
the direction perpendicular to the layers,
in other words, as expected from eliminating the interlayer hopping
interactions. 
However, the QS$GW$ gap increase signficantly and, moreover, was found
to converge very slowly with the size of the vacuum region, separating
the monolayers in the periodic boundary condition method used, even
for 2D systems. In fact, the gap then converges as $1/L$ with $L$ the
size of the vacuum region and a gap as large as 7.66 eV was obtained
by extrapolating to $L\rightarrow\infty$. Very recently, an increase in
optical band gap measured by the Tauc method
was reported for bilayer V$_2$O$_5$  compared to bulk by 1.33 eV.\cite{Reshma21}
This is larger than predicted by LDA but still significantly smaller than
predicted by QS$GW$. 

Clearly, this presents a challenging problem to our understanding of the
2D materials electronic structure  because even for the bulk, there
is a large discrepancy. From studies of other 2D materials,\cite{Komsa12}
it is already well known that such a large change of the gap in monolayers compared to bulk results from the long-range character of the self-energy
$\Sigma({\bf r},{\bf r}^\prime,E)$ in insulators and again
from the strong changes in screeening in 2D \cite{Cudazzo11}
that were already
mentioned above in the context of phonons. In other words, this leads to
an increase in $W$.  However, it should be realized
that these 2D effects relate to the quasiparticle gap or fundamental
gap $E_g^{qp}=I-A$, defined as the difference between the
ionization potential $I=E(N-1)-E(N)$ and electron affinity $A=E(N)-E(N+1)$.
This type of gap is in principle measured by taking the difference
between inverse photoemission (IPES), or
Bremsthralung isochromat spectroscopy (BIS)
and photoemission spectroscopy (PES)
onsets and measures one-particle addition or removal
energies. These are the excitation energies considered by many-body-theory
when creating or destroying a particle from the many-electron system.
However, in optical absorption, one creates a two-particle
excitation or an electron-hole pair and the interaction between the two,
which is the excitonic effect can lower the excitation energy. 
When including the two particle or absorption gap using the BSE equation,
one finds that $W$ also affects the exciton binding energy and hence
the increase in $W$ leading to a large enhancement of the quasiparticle gap
is largely compensated by the increase in exciton binding energy so that
the optical gap of  the monolayer stays close to that of the bulk. 

However, clearly in oxides such as V$_2$O$_5$, we need to  first
understand the origin of the QS$GW$ gap overestimate in the bulk
before we can attempt to understand the monolayer. So, first,
inverse photoemission and photoemission experiments by
Meyer \etal\cite{Meyer11}
show that the quasiparticle gap is about 2.8$\pm0.5$ eV.
This is substantially
higher than the optical absorption onset of 2.35 eV but still far from
the QS$GW$ gap. 
Two main reasons
come to mind for the origin of this overestimate, the underscreening
by the RPA or lack of electron-hole interactions or electron-phonon
coupling band gap renormalization effects. 

In Ref. \onlinecite{Bhandari15} we pursued the hypothesis that electron-phonon
coupling would be responsible for the discrepancy. In fact, 
Botti and Marques \cite{BottiMarques} had proposed that for polar materials
one should
take the ionic displacements or relaxation into account in the screening
of $W$ in the long-wavelength limit.  They called this a lattice polarization
correction (LPC). 
In a material with strong LO-TO splittings
this substantially increases the static dielectric constant compared to the
high-frequency (electron only screening) dielectric constant
because of the generalized Lyddane-Sachs-Teller equation
\begin{equation}
\frac{\varepsilon_0({\bf q}\rightarrow0)}{\varepsilon_\infty({\bf q}\rightarrow0)}=\prod_m \left(\frac{\omega_{LO}^m}{\omega_{TO}^m}\right)^2,
\end{equation}
which relates the dielectric screening increase to the vibrational modes $m$.
We thus
used a correction factor $\alpha=\varepsilon_\infty/\varepsilon_0$
for $W$ and hence $\Sigma$, where we averaged in some way over the
different directions. This led to an estimate of a correction factor
of 0.38 which then brought the optical absorption gap  down to 2.6 eV
which seemed reasonable.

However, this assumed that the correction factor to apply for all
wavevectors whereas according to the Botti-Marques approach it
should only be applied in the ${\bf q}\rightarrow0$ limit.
Now in the $GW$ method the self-energy evaluation in reciprocal
space  and energy rather than time domain,
requires a convolution between $W({\bf q},\omega)$ and
the Green's function $G({\bf q},\omega)$ and the singularity of $W({\bf q},\omega)$ for ${\bf q}\rightarrow0$ requires a special treatment of the
neighborhood of ${\bf q}=0$. This is done by the off-set $\Gamma$ method
described in Refs.\onlinecite{Kotani07,Kotanijpsj,Friedrich10}.
Thus by replacing the macroscopic electronic-only dielectric
constant by the one including lattice polarization for just
${\bf q}=0$, one approximately takes into account this LPC.
However, the correction then depends on how large a region around
${\bf q}=0$ this correction represents, in other words on the
integration mesh. For V$_2$O$_5$ this was reported in
Bhandari's thesis\cite{Bhandarithesis} and showed that using
a well-converged  {\bf k}-mesh for which the QS$GW$ method is converged,
the correction amounted only to $-0.15$ eV for the indirect gap and
$-0.21$ eV for the lowest direct gap.  So, this effect cannot explain
the discrepancy with experiment.  Furthermore, this could still
overestimate the correction. In fact, to properly apply the Botti-Marques
LPC, one needs to apply it in {\bf q}-space over a region set by
the inverse of the polaron length scale $a_P=\sqrt{\hbar/2\omega_{LO}m^*}$
which also involves the effective mass of the corresponding band edge.\cite{Lambrecht17} This then amounts to the classical Fr\"ohlich polaronic zero-point-motion correction apart from a factor two which arises in our model
from the fact that we limited the integration of the polaron correction
to $1/a_P$ whereas traditionally it is extended to infinity. This factor
two is still under dispute, for example a recent paper evaluated
the Fr\"ohlich contribution more completely and arrives
at estimates of this zero-point motion correction to the gap
about a factor two larger than ours for various
materials.\cite{Miglio2020}

So, at any rate, electron-phonon coupling seems not to be the primary
reason for the gap overestimate in V$_2$O$_5$ by QS$GW$. Although polaron
formation may be expected to occur given the vary flat bands, this
would primarily affect luminescence but not optical absorption and
the polaron takes some time to form.  We thus need to return to the
electron-hole interactions and the BSE approach.
Unfortunately, evaluating
the electron-hole effects on $W$ has still not been possible for V$_2$O$_5$
because of the demanding nature of these calculations for a system
with as many as 14 atoms per unit cell. 
\begin{figure}
  
  \includegraphics[width=8cm]{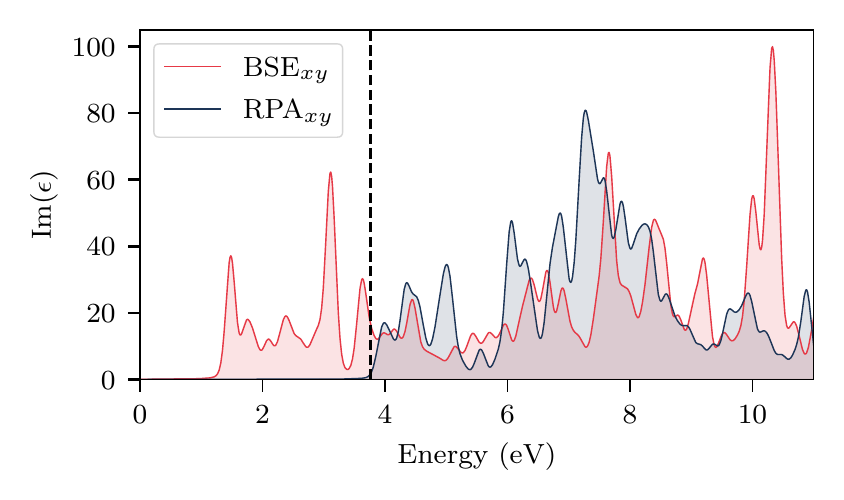}
  \caption{Imaginary part of the dielectric function of LiCoO$_2$
    calculated at the RPA and BSE levels based on QSGW($\Gamma$) \label{fig:lcobse}}
\end{figure}

Meanwhile we found that similar problems plague other oxides,
notably LiCoO$_2$. The GGA gap in this case is found to be 0.87 eV while
QS$GW$ gives a gap of 4.125 eV. Now in this case, the smaller unit cell
size allowed us to fully apply the recently developed BSE approach by
Cunningham \etal \cite{Cunningham18}. By that we mean that the BSE is
applied not only to the calculation of the  macroscopic optical dielectric
function, in other words in the limit ${\bf q}\rightarrow0$ but at all
${\bf q}$ in the Brillouin zone. This allows us,  first  of all, to re-evalute
  the QS$GW$ gap with a modified $W$. We call this QS$GW(\Gamma)$ because
  it is equivalent to including a vertex ($\Gamma$) correction
  to $W$. This yields a quasiparticle gap of 3.76 eV. However, examining 
  now the dielectric function in the ${\bf q}\rightarrow0$ limit,
  we find that the lowest peaks in absorption are excitonic in nature
  and yield an exciton gap of only 1.4 eV in agreement with
  the best experimental determinations of the absorption onset (Fig. \ref{fig:lcobse}).
  Further details of this recent work can be found in
  Radha \etal\cite{Santosh_lcooptic}
  This clearly shows that in LiCoO$_2$ the optical absorption onset corresponds
  to strong Frenkel excitons and is significantly different from the
  quasiparticle gap. The exciton binding energy is of order 2.3 eV.
  
  With this new insight, we expect that strong excitonic effects will also
  be important in V$_2$O$_5$.  In terms of the bulk monolayer gap difference,
  we then expect, that 
  similar to MoS$_2$ and other
  TDC materials, the monolayer effects on the optical gap will
  be moderate. As observed for those materials, the increase in $W$ due
  to reduced screening affects both the quasiparticle gap and the exciton
  binding energy in a similar fashion so that ultimately they largely 
  compensate each other. 

\begin{figure*}
  \includegraphics[width=16cm]{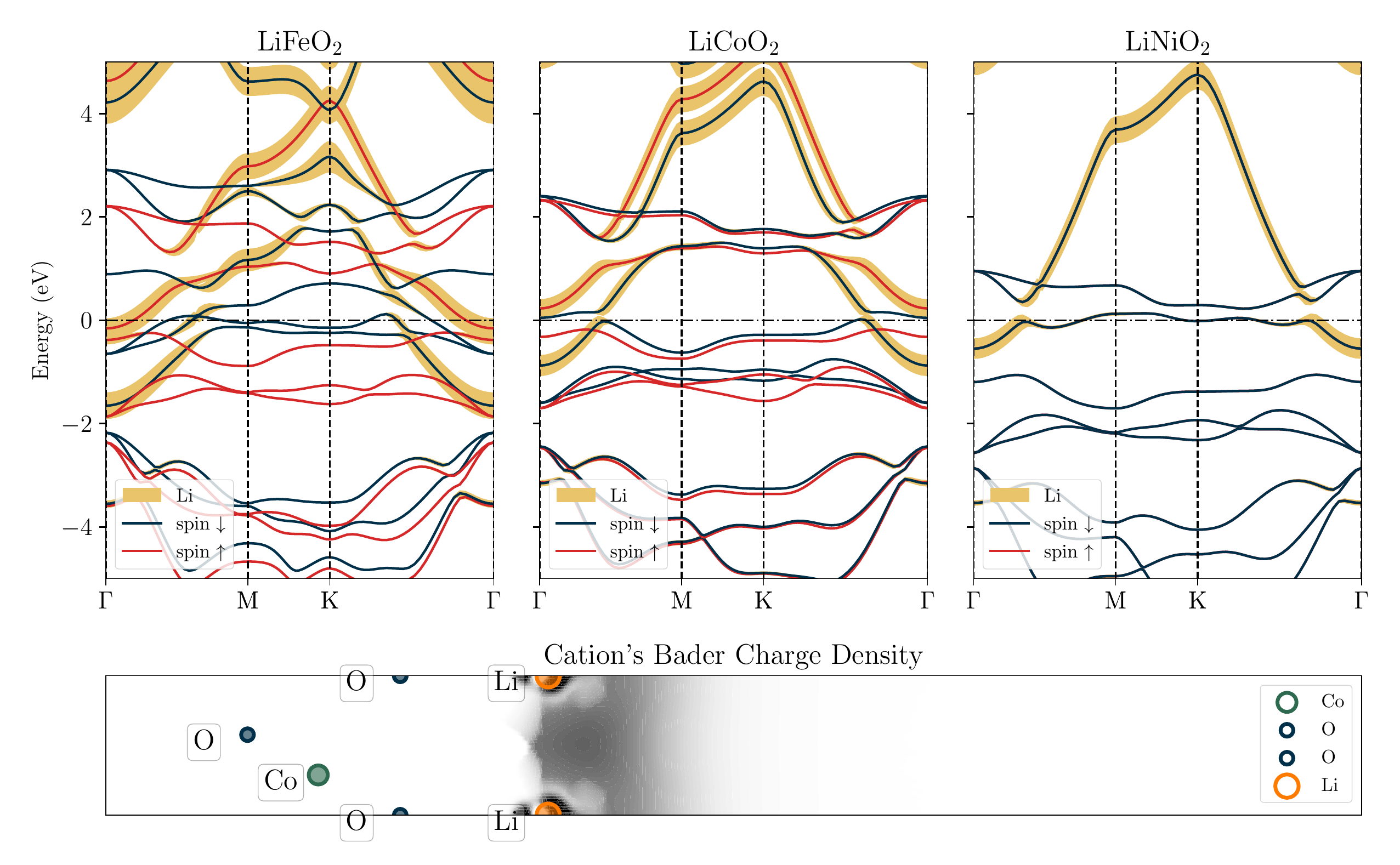}
  \caption{Comparison of LiCoO$_2$ with  NiFeO$_2$  and LiNiO$_2$
    monolayers of the $R\bar{3}m$ structure. The yellow shading
    highlights the Li related surface states which are a  result
    of the topoligical non-triviality. In both Fe and Co cases,
    the resulting monolayer band structure is spin-polarized (as indicated
    by the difference between red and black lines) while in LiNiO$_2$
    it is not.\label{figotherr3m}}
\end{figure*}
\subsection{Topological Surface States in LiCoO$_2$} \label{sec:lco2deg}
Thus far we have focused on QS$GW$ effects on the band gap of 2D materials
related mostly to the different nature of screening in 2D. However,
in some materials, 2D effects can be more profound in altering the electronic
structure even at the much simpler semilocal density approximaton level.
In topologically non-trivial materials the bulk-boundary connection
may lead to the existence of topologically protected surface states
which are partially filled and hence metallic. 
Needless to say for an ultrathin few-layer system, such effects
are predominant because they are almost all surface.

Unexpectedly, we found this to be the case in LiCoO$_2$, although
this material has not been previously recognized to be topolocially
non-trivial. This is because the notion of topology has recently
been signficantly expanded when considering crystalline symmetries
instead of only continuous symmetries such as time reversal, or particle-hole
symmetry.  While such crystalline topoligical effects are only weakly protected,
we still found them to have profound effects in the case of
LiCoO$_2$ in the $R\bar{3}m$ structure.

We discovered these effects somewhat serendipitously by performing
DFT calculations for a monolayer of LiCoO$_2$, inspired by
the exfoliation experiments.\cite{Radha2021}
In other words, we considered a  monolayer
of CoO$_2$ consisting of the same  connected CoO$_6$ octahedra as in
the $R\bar{3}m$ structure and
with Li attached to one side.  Intriguingly what we found was
that the Li-$sp_z$ orbital related bands, which in bulk occur at high energy
above the Fermi level, came down in energy and near $\Gamma$ formed
an occupied pocket below the Fermi level, which manifests itself in
the formation of a two-dimensional electron gas (2DEG) with an
increased electron density on the Li and floating above it, extending
into the vacuum region. Even more interesting, this electron gas
was found to be spin-polarized and accompanied by a corresponding hole
gas on the CoO$_2$ layer. Further calculations  of systems consisting
of several layers and replacing Li by Na and in various
surface locations confirmed the robustness of the effect.  The surprising
fact about this result is that the bonding in LiCoO$_2$ is usually
assumed to be purely electrostatic with Li levels being so high
that they donate their electron entirely to the CoO$_2$ layers. So, why
would the Li levels come down in energy at the surface? The answer is
that the covalent bonding between the Li hybridized $sp_z$ orbitals
and  the two CoO$_2$ layers above and below it in the bulk phase is non-trivial
in character and plays an important role. 

We found that it can be related to the bonding in
the quadripartite Su-Schrieffer-Heeger (SSH4) 1D chain model. To gain
further insight, we modeled the CoO$_2$ layer as just two sites with
$s$ orbitals. Within this model, if the square of the
interacton between Li and Co
is stronger than the product of the interactions between Li-Li
and Co-Co, the system is in a non-trivial topology, characterized
by a non-zero topological winding number. This implies that if the Li-Co
bond is broken at the edge, or, alternatively phrased, if the unit cell
is chosen so that  one of the Li-Co interactions is between one cell
and the next, then a metallic  (\ie half filled)
edge state is required. However, while the
original SSH4 model has chiral or particle-hole symmetry, the energy
difference between the on-site energies of Li and Co produces a gap in this
edge state and would still transfer the electrons to the lower energy
edge, in this case the CoO$_2$ layer. In other words, it is no longer
topologically protected. 
The key to why nevertheless
some electrons remained with the Li lies in  the lateral interactions
parallel to the surface. These are stronger for the Li than for
the CoO$_2$ and hence the band broadening of the Li surface band
allows it to overlap with the CoO$_2$ surface  derived surface band
creating a semimetallic situation with partial electron occupation
on the Li surface and hole occupation on the CoO$_2$ surface. Crucial
for this to be enabled is that the Li
surface concentration is sufficiently high to allow for sufficient band
width. 
Note that while the Li surface band center lies
higher in energy than the CoO$_2$ one, it still is   significantly
lower in energy than in the bulk because the topological effect
brought it down. The spin-polarization in this case results from
the holes present in the CoO$_2$ layer, which is equivalent to
reduced Li concentration in Li$_x$CoO$_2$ and, here, because of the proximity,
also polarizes the Li-side 2DEG. 
The details of this study can be found in Radha and Lambrecht
\cite{Radha2021}

The topological effect discussed above is not limited to LiCoO$_2$
but occurs in other oxides with the $R\bar{3}m$ structure.
Fig. \ref{figotherr3m}
shows  the Li related surface band in LiCoO$_2$, LiFeO$_2$ and LiNiO$_2$
monolayers. The bottom part of Fig. \ref{figotherr3m} shows the charge density
associated with Li in a LiCoO$_2$ monolayer using the Bader charge method. 

The impact of this topology related surface 2DEG formation goes beyond
the isolated 2D monolayer. We found subsequently that as the
distance between CoO$_2$ layers is expanded in a periodic system,
there is a critical transition where Li instead of staying at equal
distance from the two neighboring CoO$_2$ layers, associates with one layer.
The system at this point transforms into a set of non-interacting monolayers.
Guided by the chemical exfoliation procedures, one may envision that this could
be achieved by inserting large organic molecules in between the layers
or by inserting H$_2$O molecules. As discussed in Sec. \ref{sec:chemexfo},
the role of water molecules in the process is not yet fully understood
but appears to be crucial to achieve successful
exfoliation. In fact, such a route to obtain essentially 2D physics
within an extended periodic system has already been used in hydrated
NaCoO$_2$. In  this  system, specifically in  Na$_x$CoO$_2$:yH$_2$O, ($x\approx0.35$, $y\approx1.3$) superconductivity has been discovered at 4K.\cite{Takada2003,Schaak2003}
Remarkably we find that the required tensile strain of the interlayer
distance in NaCoO$_2$ where the symmetry breaking
transition to isolated monolayers occurs, is much lower in the Na
than the Li case and coincides closely with the distance obtained
in the hydrated layers were superconductivity occurs. We thus speculate
that the superconductivity may be related to the presence of either the
electron or hole electron gases in this system, which has thus far
not been fully appreciated.

\section{Transport in 2D Systems} \label{sec:transport}
\subsection{Tranport in V$_2$O$_5$}
\subsubsection{Highly Anisotropic In-Plane Transport in $\alpha$-V$_2$O$_5$} \label{sec:v2o5transport}

\begin{figure}
	\centering
	\includegraphics[width=0.48\textwidth]{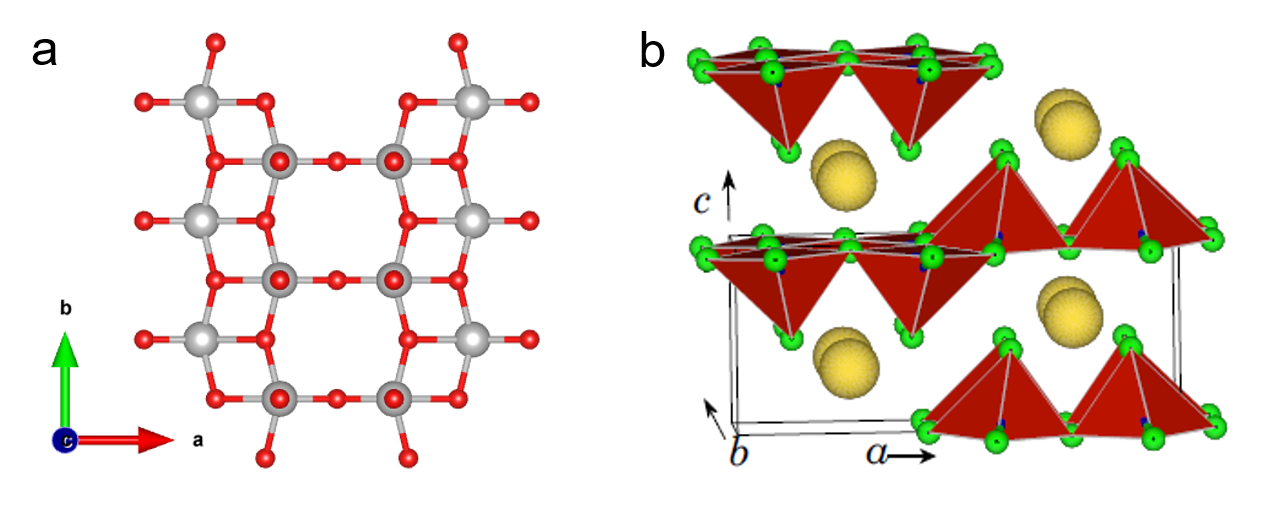}
	\caption{a) Lattice diagram of V$_2$O$_5$. b) Lattice diagram of NaV$_2$O$_5$. Figure adapted from Bhandari et al.\cite{Churnanav2o5} with permission.}
	\label{fig:v2o5}
\end{figure}

The family of V-O compounds contains many species, due to vanadium’s wide range of oxidation states from 2+ to 5+.\cite{Zeiger1975, Eyert2011, Beke2011} Numerous forms have found use in a range of applications, such as chemical sensors, anti-reflection coatings, and oxidation catalysts.\cite{Fiermans1980, Gopinath2000, Velusamy2003, Jiang2007, Micocci1997} V$_2$O$_5$, the most stable member of the family, exists in a variety of phases. In this perspective, we are concerned with the $\alpha$-phase, with all references on V$_2$O$_5$ from this point pertaining to this phase. V$_2$O$_5$ is a well-studied orthorhombic van der Waals “ladder” compound of space group $Pmmn$.\cite{Ou-Yang2012} The reason for the name is apparent from figure \ref{fig:v2o5}a, where connected double zigzag chains of V-O propagate in the {\bf b} direction and are connected by “rungs”
via a bridge oxygen in the {\bf a} direction. Furthermore, it turns out that in NaV$_2$O$_5$ the electrons located on
these V-O-V rungs in the lowest split-off conduction band with V-$d_{xy}$ orbital character have alternating spin orientation
along the ladder direction and play a key role in its magnetic properties. 
There is also sufficient spacing in between layers for ionic intercalation with group-I and group-II elements, as demonstrated by figure \ref{fig:v2o5}b for sodium. 

Much like MoO$_3$, the conductivity of V$_2$O$_5$ is largely determined by the atomic composition and stoichiometry of the crystal. In its pristine state, V$_2$O$_5$ is an insulator, becoming natively conductive only when oxygen is removed. The lattice consists of single, double, and triply coordinated oxygen species, of which the singly coordinated oxygen is most likely to leave in reducing conditions.\cite{Smolinski1998} As oxygen bonds are broken, electrons are donated to the lowest conduction band, which is called the “split-off” band in V$_2$O$_5$, increasing n-type conductivity and mobility.\cite{Scanlon2008} 

The mobility increase in V$_2$O$_5$ is surprisingly anisotropic. Due to the
larger orbital overlap in the V-O chains in the {\bf b}-direction than in the {\bf a}-direction, electrical conduction is greatly enhanced, resulting in a strongly 1D transport.\cite{Sucharitakul2017} Similarly, phonon-polariton
transport in V$_2$O$_5$ has also been found to be highly anisotropic.\cite{Taboada2020}

Recently, exfoliation experiments on V$_2$O$_5$ crystals were conducted by Sucharitakul \textit{et al.} \cite{Sucharitakul2017} Interestingly, the mechanical exfoliation of these crystals led mostly to rectangular nanoflakes with large aspect ratios.  This is related to the presence of the above mentioned 1D chains within the layer. This means that the short side of each flake is expected to be perpendicular to these chains because these chains are connected by a bridge oxygen while the oxygen within the chain is bonded to three vanadiums. (See Fig. \ref{fig:v2o5}.) Confirmation of this was obtained by performing Raman spectroscopy.  The A$_{1g}$ modes near 482 cm$^{-1}$ have a strong anisotropy in their Raman intensity which allowed us to identify the chain direction of the crystal relative to the long or short axis of the nanoflakes.  Flakes with thicknesses between 10 nm and 150 nm were obtained, which is still far from a monolayer system.  The c-axis of V$_2$O$_5$ perpendicular to the layers is only 4.8 \AA, so even the smallest flakes are still approximately 20 layers thick.  Contacts were made to flakes of $\mu$m lateral size and showed a marked anisotropy of the conductivity as measured by four probe van der Pauw method Hall measurements in addition to measuring sheet conductances.  The ratio of the conductivity along the chain versus perpendicular to the chain was found to be as high as a factor of 10.  Estimates of the conductivity and conductivity ratio were performed within a model based on in-plane acoustic phonon scattering using an approach previously used for MoO$_3$\cite{Zhang2017} and gave a satisfactory explanation of this anisotropy ratio.  The large anisotropy of conductivity within the layers has previously been reported for bulk-sized samples\cite{Yosikawa1997} but was studied based on a variable range hopping conduction mechanism.  However, the calculated conductivities were of order 306 cm$^2$/(Vs) along the chains while actual measurements gave much lower conductivities of order 7 cm$^2$/(Vs).  The calculated theoretical limits on the mobility were comparable to those reported for MoO$_3$.  This indicates that there is much room for improvement of the mobilities in nanoflake V$_2$O$_5$.

In V$_2$O$_5$ the conductivity is n-type and corresponds to filling of the bottom of the split-off conduction band, which primarily has dispersion along the chain direction.  Nonetheless, it is not only the lower mass in this direction that plays a role in the conductivity but also the anisotropy in the deformation potentials and elastic constants in the plane. The origin of the carriers could be arising from the oxygen vacancies, such as vanadyl oxygen vacancies or could arise from intercalating impurities, such as Li and Na.  The limitations in carrier concentration in V$_2$O$_5$ used in these nanoflake studies were limiting the measurements and could be overcome by intentional doping or intercalation in future studies.

\subsubsection{P-type transport and Low Dimensional Scattering in $\alpha$-Na$_{0.96}$V$_2$O$_5$}

Another way in which the conductivity of V$_2$O$_5$ may be impacted is via ionic intercalation, as in Fig. \ref{fig:v2o5}b. This produces what are known as bronze phases, with documented studies for species such as Li, Cs, Na, K, among others.\cite{Taboada2020, Ueda1998, Mumme1971}
Li intecalation in
V$_2$O$_5$ has been extensively studied in the context of Li-battery cathodes but leads to somewhat larger distortions of the lattice than Na intercalation, 
related to formation of the $\gamma$-form\cite{Willinger04} 
of V$_2$O$_5$.\cite{Jarry20,Walker20,DeJesus2016} Mg-intercalated V$_2$O$_5$ in the $\beta$-form\cite{Filonenko04}
has also recently been considered as a battery material. 

\begin{figure}
	\centering
	\includegraphics[width=0.48\textwidth]{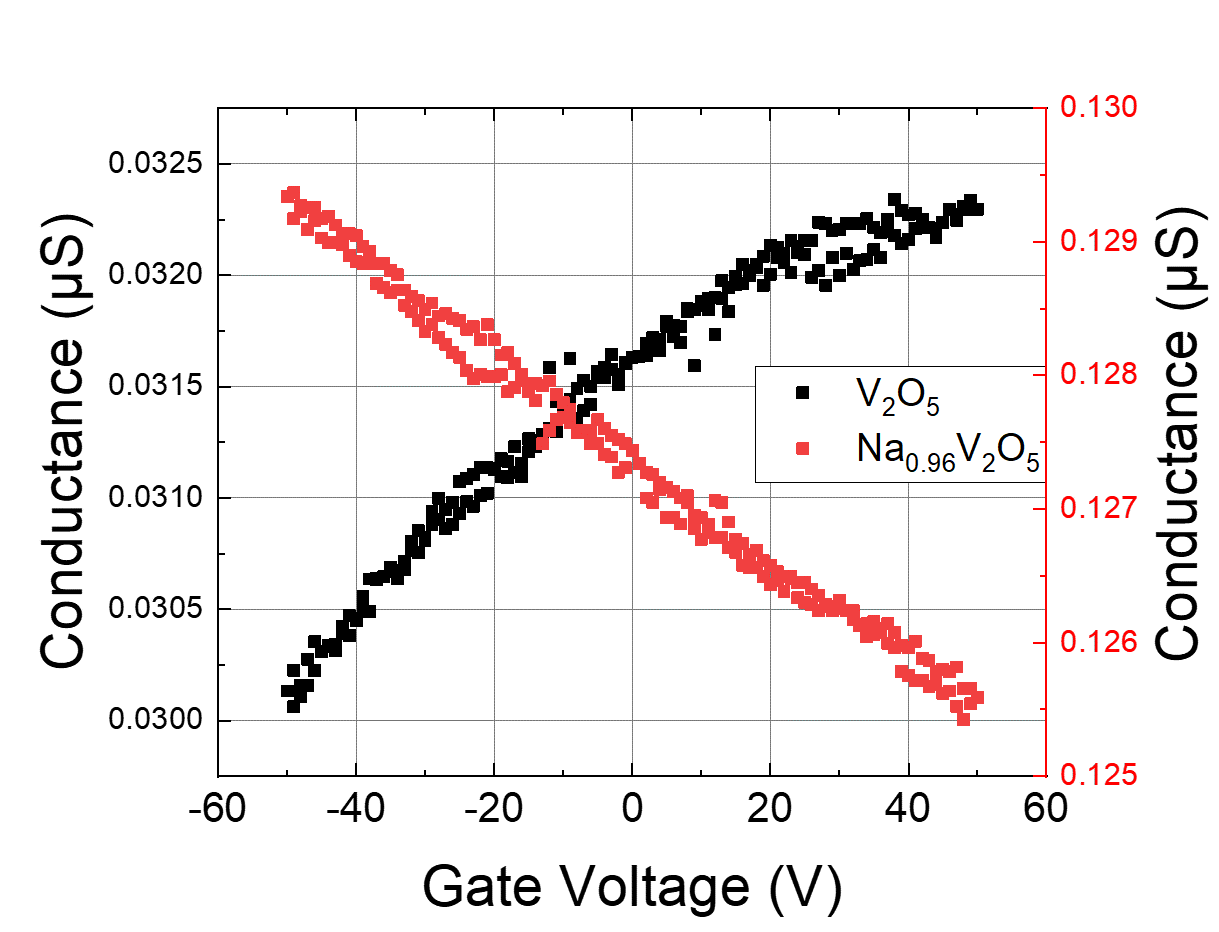}
	\caption{Field-effect gate response for V$_2$O$_5$ and Na$_{0.96}$V$_2$O$_5$.
	\label{fig:mobility}}
\end{figure}

In this perspective, we pay attention to sodium doping, which has been utilized to influence magnetic ordering and optical phonon polariton tuning. The ionic radius of sodium is small enough that it minimally impacts the lateral spacing of the crystal, only increasing the interlayer spacing in the stacking direction, while still donating electrons to the split-off band.\cite{Kanke1990,Enjalbert1986,Churnanav2o5} For this reason, sodium intercalation is a preferential candidate to examine changes to electronic transport without significantly affecting crystal symmetry.
Recently, this dopant was investigated by our collaboration. P-type conduction was observed in few-layer flakes of Na$_{0.96}$V$_2$O$_5$. Using the scotch-tape mechanical exfoliation method, few-layer flakes were isolated, followed by subsequent deposition of electrical contacts in a 2-probe FET configuration. Similar thickness flakes of undoped V$_2$O$_5$ were also characterized simultaneously for comparison, as in Figure \ref{fig:mobility}a. Positive/negative transconductance slopes for V$_2$O$_5$/Na$_{0.96}$V$_2$O$_5$ were observed, indicating a majority carrier of electrons/holes, respectively. Performing a linear regression, carrier mobility can be extracted. During our investigation, sodium-doped V$_2$O$_5$ mobility was observed to diminish in magnitude as flake thickness decreased, in contrast to our observations for undoped V$_2$O$_5$ flakes across the same thickness range. While a more thorough investigation of this behavior is required, we offer the speculation that sodium presence, or possibly absence, at the surface of the exfoliated Na$_{0.96}$V$_2$O$_5$ flakes may play a role, locally scattering charge carriers and inhibiting conduction in thinner system.

Surprisingly, of the experimental characterization which has been performed on electrical transport in NaV$_2$O$_5$ systems in the past, there is only a single mention of p-type conductivity by Carpy \textit{et al.}\cite{Carpy1972} who
obtained this result from thermoelectric measurement.  However, the subject does not appear to have received attention as a focus in an experimental transport capacity as of yet.\cite{Carpy1972,Chakrabarty1976,McNulty2019,Lohmann1999}
Given the contact resistance that accompanies a 2-probe configuration, the mobility extracted in our study of both systems is a few orders smaller than was
observed in Sucharitakul \textit{et al.} \cite{Sucharitakul2017}
 Even so, we expect that this p-type mobility in NaV$_2$O$_5$ may be engineered to more significant magnitudes by controlling sodium stoichiometry. 

 Whereas oxygen deficiency in V$_2$O$_5$ donates electrons to the split-off band and leads to n-type conduction, sodium intercalation, with one Na ion per V$_2$O$_5$, leads to a splitting of the split-off band into spin-up and spin-down components, where charge donated by sodium fills the lower of these two bands. This half filling is directly linked to antiferromagnetic ordering observed in NaV$_2$O$_5$, and is examined in detail by Lambrecht and Bhandari.\cite{Lambrecht81,Churnanav2o5} In the case of partial doping (that is, for Na$_x$V$_2$O$_5$, $x < 1.0$), the filling of the lower split-off band is incomplete, and results in a small fraction of empty acceptor states, leading to p-type transport. 
It follows that a greater deficit of Na should result in a greater p-type conduction, which should be limited only by the instability of the $\alpha$-phase of Na$_x$V$_2$O$_5$ beyond $x < 0.7$.\cite{Kanke1990} Given the novelty of p-type transition metal oxides and their utility in the semiconducting industry, continued investigation of Na$_x$V$_2$O$_5$ at various stages of intercalation may lead to fruitful applications of this compound.

\subsection{Transport in MoO$_3$}
\subsubsection{Doping and Transport in MoO$_3$}

Transition metal oxides (TMOs) are often overlooked as candidates for the conducting component in 2D transistor applications because of  their large native insulating behavior.  However, these oxides offer a wide range of applications
after minimal modification. MoO$_3$ has demonstrated its usefulness as a multi-functional material with its current uses ranging as an electrochromic material\cite{J.N.Yao1998,Faughnan1977}, an optical and gas sensor\cite{Rahmani2010,Comini2005}, and a hole-injection layer for organic photovoltaic cells.\cite{Girotto2011,Hancox2010} Recently MoO$_3$ has also been investigated for its applications in 2D field effect transistors (FETs) as both the conducting channel\cite{Peelaers2017,Crowley2018} and as the gate oxide dielectric material.\cite{Holler2020}

As mentioned in Sec. \ref{sec:intro}, MoO$_3$ has an orthorhombic $\alpha$-phase which allows for simple mechanical exfoliation.  Specifically, $\alpha$-MoO$_3$ is composed of double-layers of distorted MoO$_6$ octahedra bonded to adjacent layers via weak van der Waals (vdW) forces, which allows for isolation and study of 2D and few-layer systems.\cite{Chithambararaj2016} 
Intrinsic $\alpha$-MoO$_3$ has a large work function of $>6.9$ eV and an even deeper ionization energy of $>9$ eV which leads to a bandgap of $\sim$3eV.\cite{Guo2014,Meyer2015}

Various methods of doping have been utilized to tune the electrical properties of $\alpha$-MoO$_3$, enabling exploration of this oxide as a FET. For example, recently Crowley \textit{et al.} demonstrated that exposure to H$_2$ gas at temperatures ranging from 275-400$^\circ$C allowed for controllable reduction to a substoichiometric $\alpha$-MoO$_{(3-x)}$.\cite{Crowley2018} Characteristics such as conductivity, field effect mobility, and carrier concentration increased as a function of reduction temperature, and showed consistent n-type behavior. Electron conduction in $\alpha$-MoO$_{(3-x)}$ is indeed consistent with theoretical predictions,\cite{Zhang2017} despite previous accounts of hole-conduction in similar systems by other groups.\cite{Balendhran2013, Alsaif2016}

Recently, the creation of oxygen vacancies via hydrogen doping has been used to purposefully tune the conductivity and electron mobility of MoO$_3$, making it a more suitable and controllable candidate for 2D transistors.\cite{Crowley2018,Balendhran13,Alsaif2016} For instance, Crowley \textit{et al.}\cite{Crowley2018} demonstrated that exposure to H$_2$ gas at temperatures ranging from 275-400$^\circ$C caused a reduction in the MoO$_3$, forming a more conductive substoichiometric MoO$_{(3-x)}$.  This study revealed that the conductivity, field effect mobility, and carrier concentration increased as a function of reduction temperature, and thus reduction rate, and showed consistent n-type behavior of the modified MoO$_3$.

Another method of transport enhancement under recent investigation is via utilization of fluorine. Fluorination of $\alpha$-MoO$_3$ has until recently only been examined using first-principles methods, where F is predicted to replace the doubly coordinated oxygen and facilitate Mg diffusion throughout the
lattice.\cite{Wan2016} Experimentally, F exposure via dry-etching with SF$_6$ plasma was found to impact n-type mobility in a similar manner as oxygen deficiency, albeit to a lesser magnitude. Interestingly, this effect was observed to be reversible upon annealing the exposed flakes in an inert atmosphere, returning the $\alpha$-MoO$_3$ to an insulating state. F presence was observed to impact the terminal oxygen in $\alpha$-MoO$_3$, suggesting that fluorine may provide a means of engineering n-type effects via surface interaction, without permanently impacting oxide stoichiometry.

\subsubsection{Opportunities for 2D MoO$_3$-based Transistors}

\begin{figure}
	\centering
	\includegraphics[width=0.48\textwidth]{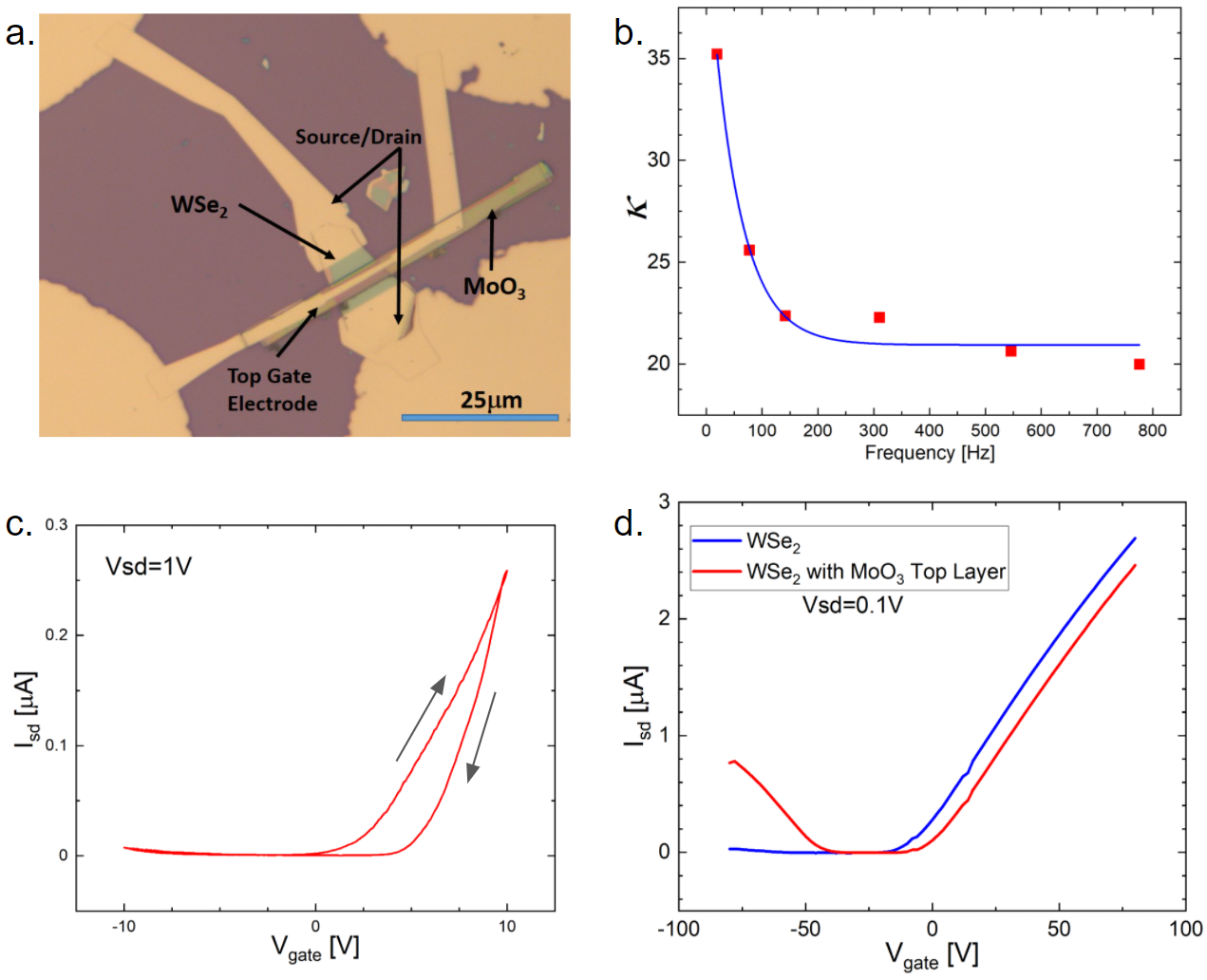}
	\caption{a) Optical photograph of WSe$_2$/MoO$_3$ FET. b) Dielectric constant, $\kappa$, of MoO$_3$ ranging over low frequencies. c) Top-gate behavior of WSe$_2$/MoO$_3$ FET. d) Hole inducing behavior after exfoliated MoO$_3$ is layered on top of WSe$_2$. Figure adapted from Holler et al.\cite{Holler2020}
with permission.}
	\label{fig:Dielectric}
\end{figure}

Since $\alpha$-MoO$_3$ intrinsically has a band gap of $\sim$3eV, it seems natural to realize its function as a dielectric material for 2D semiconductor devices.  In a recent paper, Holler \textit{et al.}\cite{Holler2020} demonstrated that MoO$_3$ can act as the top-gating material in layered 2D semiconductor FETs. This was motivated by experimental claims of thin-film MoO$_3$ having a large dielectric constant.\cite{Naouel2011a,Saad2005,Sayer1978,Ren2017b} some reporting a low-frequency $\kappa$ as high as $\sim$200.\cite{Naouel2011a}
In their study, they fabricated MoO$_3$ single crystal parallel-plate capacitors and measured the 90 degrees out-of-phase ac current to calculate a dielectric constant $\kappa$ approximately 35 near 0Hz, (figure \ref{fig:Dielectric}a), which is approximately ten times greater than that of SiO$_2$ and also greater than other comparable non-vdW high-$\kappa$ oxides such as ZrO$_2$ and HfO$_2$.\cite{Liu2018,Robertson2006,Manchanda2001,Park2012} This high value of the static dielectric constant differs significantly from the calculated values\cite{Amol2020}
for pure insulating $\alpha$-MoO$_3$ including phonon contributions, which are 27, 13 and 7 along $b$, $c$, $a$ in the standard $Pnma$ setting
of the spacegroup.  This indicates an extrinsic origin related to the non-zero
conductivity in these samples.

WSe$_2$/MoO$_3$ heterostructures were fabricated by exfoliation via the adhesive tape method and then transferred onto a 300nm Si/Si$^{++}$ substrate.  The MoO$_3$ was carefully stacked onto the WSe$_2$ via a polydimethylsiloxane (PDMS) stamp.  The source and drain electrodes in addition to the top-gate were fabricated via electron-beam lithography, as seen in figure \ref{fig:Dielectric}b.  As presented in figure \ref{fig:Dielectric}c, MoO$_3$ proves to be capable of properly gating the WSe$_2$ device, sweeping the gate voltage from -10V to +10V.  The electron field effect mobility, $\mu_{\textrm{FET}}$, was calculated to be 1.62 cm$^2$V$^{-1}$s$^{-1}$ and can be improved by increasing the surface area coverage of of the top gate material with respect to uncovered WSe$_2$.  The device exhibits an on/off ratio of approximately 10$^3$ and a subthreshold swing of approximately 2.2V/dec.

Due to its deep work function and relative band alignment, the MoO$_3$ top-gate was also shown to induce holes in the WSe$_2$ layer, allowing for another option to control and modify the properties of the  conducting channel.\cite{Meyer2012,Peng2016}  This comparison was performed using the SiO$_2$ back-gate before and after the MoO$_3$ layer was stacked.  The device illustrates a threshold shift $\Delta V_{\rm T}=+7$V, indicating a shift towards more p-type behavior, in addition to a clear p-type response at negative gate voltages (see figure \ref{fig:Dielectric}d).

MoO$_3$ has proven itself to be an interesting candidate for many 2D transistor applications.  It has been shown that hydrogen doping of MoO$_3$ consistently resulted in n-type semiconducting characteristics, and it was observed that MoO$_3$ reduces much more readily than previously reported.  This was confirmed via measurements of conductivity, mobility, and carrier concentration for multiple annealing temperatures in a hydrogen forming gas.
Also, the effects of fluorine doping were observed with MoO$_3$: it has been shown that F presence likely impacts certain oxygen species, such as replacement of the doubly coordinated oxygen, or can adsorb on the  surface.
Lastly, in contrast to studying its conductive properties, intrinsic, undoped MoO$_3$ displays a large dielectric constant at DC and low-frequency measurements and $\alpha$-phase MoO$_3$ has been demonstrated to work as the gate oxide dielectric material for 2D semiconductor FETs.  The culmination of these studies have validated that MoO$_3$ is worth continued research due to its expressive multi-functionality and potential applications.  

\subsection{Transport in 2D LiCoO$_2$ Nanosystems}
LiCoO$_2$ (LCO) belongs to the rhombohedral space group R$\overline{3}$m with trigonal layer symmetry, with alternately stacked LiO$_6$ and CoO$_6$ octahedra in the c-direction.\cite{Levasseur2002, Orman1984} Typically, this material is characterized in its bulk form ($>0.1$mm thickness) as single crystals or compressed polycrystalline pellets, due to the strong cohesion between layers from electrostatic stabilization. LCO is often studied as a processed powder, or obtained through ion replacement methods from NaCoO$_2$.\cite{Miyoshi2015} Delithiation studies use chemical or electrochemical (EC) methods\cite{Hertz2008, Mohanty2011, Imanishi1999, Motohashi2009} to examine the effects of partial lithium loss.
As already mentioned in Sec. \ref{sec:intro}, in its stoichiometric form, LCO is insulating, where all electrons are paired and cobalt is in the low-spin configuration of Co$^{3+}$.\cite{Mizokawa2013} As the oxide is delithiated, it loses an electron with each lithium ion, forcing the surrounding lattice to compensate the charge loss. While this is often referred to as the formation of Co$^{4+}$
ions, it is not evident whether localized Co$^{4+}$ ions
actually form or whether the holes are distributed in extended states as in a p-type semiconductor. Wolverton and Zunger \cite{Wolverton98}
noted that the net charge in a sphere around
Co actually barely changes upon delithiation
and the different nominal valence is rather
accomodated by changes in the degree of covalency of the O-Co bonds. 
As mentioned in Sec. \ref{sec:anneal}
the highest average Co valence deduced from
electron loss spectroscopy of the $L_2/L_3$ edges is only 3.3 for
Li$_{0.37}$CoO$_2$.\cite{Volkova21}
 Thus, delithiation in Li$_x$CoO$_2$ leads to both increased p-type conductivity via thermally activated hopping conduction\cite{Ishida2010, Menetrier1999} as well as the emergence of an overall magnetic moment.\cite{Hertz2008, Ou-Yang2012, Miyoshi2010} As delithiation continues, conduction transitions from an insulating variable-range hopping to a classical metallic conduction, with a transition typically observed around the $x = 0.75-0.9$ range.\cite{Ishida2010,Miyoshi2010,Miyoshi2018,Molenda1989} However, a low carrier concentration, coupled with a variety of scattering mechanisms (such as lithium vacancies) results in low mobility and a small magnitude of current.\cite{Menetrier1999}

 The transition of LCO to a metallic phase is also accompanied by the phenomenon known as charge ordering, which tends to occur in strongly correlated oxides, such as magnetite and doped magnetite.\cite{Savitzky2017,Senn2012}
The commonly accepted mechanism of charge ordering is such that there is a repeating alternation of oxidation states among the metal cation species in an ionic lattice, occurring at low enough temperatures where charge fluctuation due to ion diffusion is no longer energetically favorable.
For LCO, as the ratio of Co$^{4+}$/Co$^{3+}$ increases, charge is balanced among the surrounding Co sites for each lithium vacancy to minimize energy, achieved via localized lithium diffusion throughout the lattice. At low enough temperature, lithium diffusion is halted, and the Co network assumes a stable charge ordering. This coordination can impact macroscopic characteristics, creating anomalies or jumps in electrical resistance and magnetic susceptibility when monitoring across the transition temperature. This has been observed in many cases for LCO, with a documented transition at T\textsubscript{s} = 150-175K, over a lithium content range of $x=$0.46-0.78.\cite{Miyoshi2015, Hertz2008, Motohashi2009, Ishida2010, Miyoshi2010, Iwaya13, Kikkawa1986, Liu2015, Mukai2007, Sugiyama2009} It has been assumed that this ordering favors fractional lithium contents of $x=2/3$ and 1/2,\cite{Miyoshi2018} with calculations of another minima in energy at
$x=1/3$, though this last fraction has not been conclusively observed.\cite{Sugiyama2009}

Despite all of the previously detailed delithiation effects of LCO, this process is actually still not well-understood. For example, while the insulator-metal transition is commonly observed in many studies, there is no consensus on an exact value of lithium content required for this change to occur, as is evident in the broad range \textit{x} reported above. Further, the charge ordering transition T$_s$ is observed over large temperature ranges and lithium concentrations, and has proven extremely difficult to predict. In a 2012 article, Yang et al. compend anomaly observations among 12 separate studies, comparing forms of LCO and the deintercalation methods used.\cite{Ou-Yang2012} Observations vary for multiple phases of LCO (compressed powder pellets, single crystals, thin films), where multiple approaches are used for preparation and delithiation (chemical, EC-galvanostatic, EC-potentiostatic, ion exchange + chemical). The takeaway is that this anomaly is observed in a variety of experimental conditions, yet at times is not present when anticipated. 

From our perspective, this ambiguity is in large part due to most previous investigation taking place on bulk systems using polycrystalline or compressed powders, instead of few-layer, single-crystal LCO. With bulk systems, galvanostatic and chemical delithiation typically occur radially inward, where outer layers delithiate to a greater extent than do inner layers. The degree of delithiation in such studies is typically assessed via measurement of surface potential, which masks the radial inhomogeneity of Li distribution.\cite{Ou-Yang2012} In such cases, there is a large degree of variability permitted, which may explain the broad range of observations from previous studies. If charge ordering effects in LCO are to be investigated with any consistency, it becomes imperative to develop a method for isolating thin single-crystal LCO flakes. The dimensionality of such a mesoscopic systems is advantageous in this pursuit, with flake areas large enough to facilitate nanofabrication, yet small enough to minimize any radial delithitaion effects.

As previously detailed in Sec.\ref{sec:chemexfo}, the methods for LCO exfoliation first developed by Kim et al. have been improved upon, investigating electrostatic bonding between the transition metal oxide and alkali ion layers.\cite{Pachuta19,Pachuta20} Chemical exfoliation of LCO powder leads to large, swelled powder granules. Once dried, these can then be mechanically exfoliated, producing 10-100nm thick single-crystal flakes of LiCoO$_2$. It is this step of following chemical exfoliation by scotch-tape mechanical exfoliation which has led to the largest lateral area single-crystal flakes of LiCoO$_2$ yet observed. The size of these flakes enables the use of 2D electrical characterization techniques, such as 4-probe Hall measurements, to be utilized for LiCoO$_2$ for the first time. Our exploratory investigation has already provided much perspective between bulk and 2D transport, as well as some experimental insight on the complex correlated effects and phenomena detailed in section 5B.

%%%%%%%%%%%%%%%%%%%% Figure/Image No: 1 starts here %%%%%%%%%%%%%%%%%%%%

\begin{figure}
	\centering
	\includegraphics[width=0.45\textwidth]{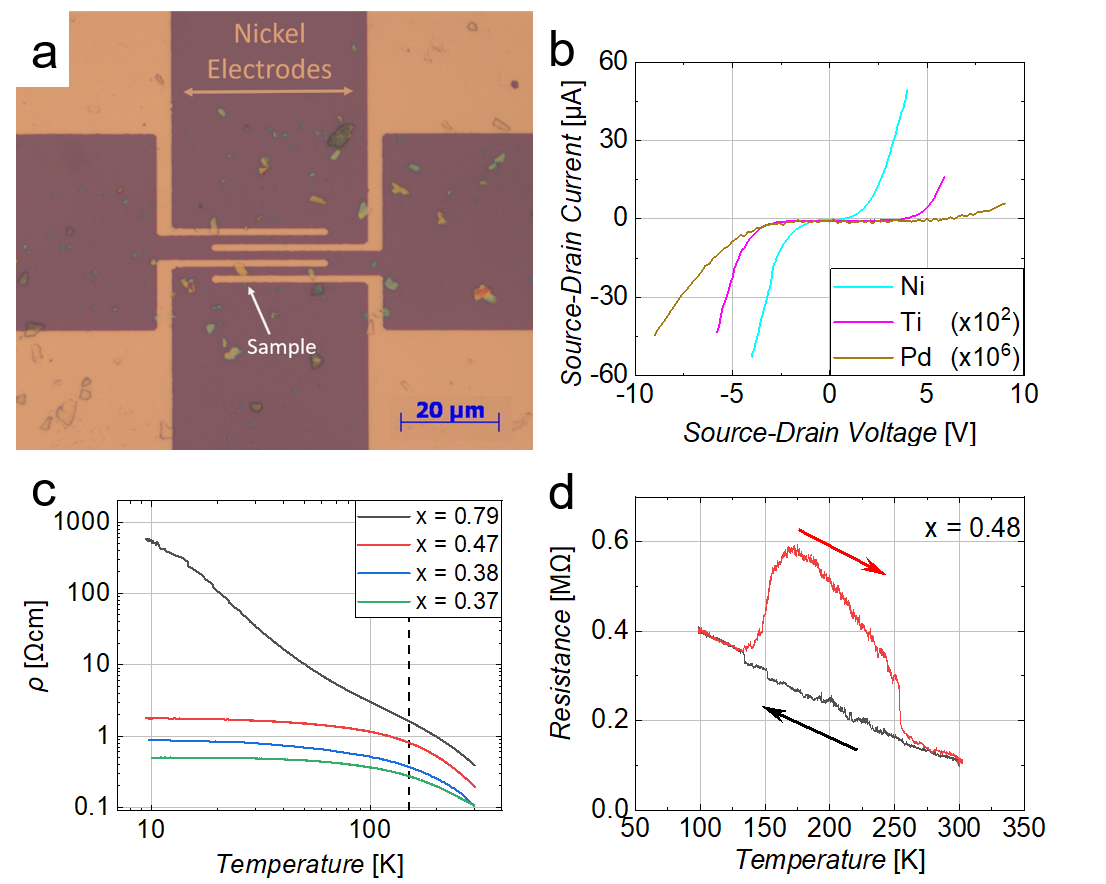}
	\caption{a) Optical image of Li\textsubscript{x}CoO$_2$ flakes with assembled source and drain electrodes. b) Source-drain curves of 2D Li\textsubscript{x}CoO$_2$ flakes using various contact metals. Current magnitudes displayed here were multiplied by a factor for each metal, indicated at bottom-right; Ni has highest magnitude of current. c) Transport in Li$_x$CoO$_2$. Various lithium concentrations are indicated at top-right. Dashed line is inset at T$_c$=150K. d) Charge ordering effects in Li$_x$CoO$_2$ flakes. Anomalies in the resistance data were observed for 3-day prepared $x$=0.46 flakes upon warming up. The anomaly begins at around T$ \sim $ 150K during the warm up. Figures adapted from Crowley et al. with permission.}
	\label{fig:LCO}
\end{figure}

%%%%%%%%%%%%%%%%%%%% Figure/Image No: 1 Ends here %%%%%%%%%%%%%%%%%%%%

In the 2D form, Li$_x$CoO$_2$ flakes are able to be masked, and metallic contacts deposited on the ends of the flake, as in Figure \ref{fig:LCO}a. Perhaps the most obvious consequence of few layer LCO is the contact energy barrier formation for all tested conditions. Under a variety of metals and materials, Li$_x$CoO$_2$ nanoflakes exhibit a Schottky barrier in their IV curve (Figure \ref{fig:LCO}b). When using divalent metals for electrodes, a doping effect with increased current magnitude is observed, seen here for the case of Ni. This is anticipated, and detailed in full elsewhere.\cite{Crowley2020, Tukamoto1997} The 2D LCO flakes of various lithium contents also reflect bulk transport characteristics, exhibiting lower resistivity as delithiation increases, and an insulator-metal transition below $x$=0.79 (figure \ref{fig:LCO}c).\cite{Crowley2020} 

Further experimentation is required to deduce the origin of the persistent Schottky barrier in Figure \ref{fig:LCO}b. However, it is interesting to note that our exfoliation method discussed in Sec.\ref{sec:chemexfo} preferentially cleaves LCO along lithium planes, potentially leaving at least a partial exposed lithium layer on the surface of the flake. While at present we have no direct
experimental confirmation of the Li termination of the surface or even less
a quatitative determination of how much Li remains on the surface, the presence
of Li on cleaved LCO single crystal surfaces has been observed by Iwaya
\etal\cite{Iwaya13}. The consequence  of a  terminal lithium layer
was the subject of discussion in Sec. \ref{sec:lco2deg} and \cite{Radha2021} where a 2DEG was predicted to exist above the surface of the flake. We offer the speculation that if the electron gas were indeed present, it would certainly pose a large energy barrier needing to be overcome for hole-conduction between LCO and the contacting metal to take place, which could explain qualitatively
the Schottky barrier behavior observed here.

As was mentioned previously, a unique advantage of using the delithiation technique presented in Sec. \ref{sec:chemexfo} is the low dimensionality of the resulting LCO flakes. In terms of electrical characterization, a bulk system may contain radial inhomogeneities of Li, spanning several hundreds of microns between electrodes, whereas a single-crystal nanoflake with a 5-10 micron conduction channel may have a higher likelihood of uniform Li distribution. Indeed, inhomogeneous Li distributions in LCO often lead to observation of a spinel phase, whereas our own nanoflakes were observed to be in the more stable  $C2/m$ phase as
discussed in Sec. \ref{sec:anneal} and \cite{Volkova21}

The concept of charge ordering relies on Li ion distributions that are orderly enough to allow fluctuation and settling into a repeating pattern at low temperature. Curiously, to this effect, anomalies in the resistance-temperature curves were observed in prepared batches of $x$=0.48 and 0.37 flakes which had been chemically delithiated over a 3 day period (figure \ref{fig:LCO}d). Samples which had delithiated at a faster rate, over 1 day with stronger acid concentration, did not possess charge ordering phenomena despite having near-identical lithium content. Additionally, the anomalies in resistance for thin LCO flakes were markedly larger in magnitude than most previous bulk reports. In one case, a repeatable $>$1M$\Omega$  increase in resistance was observed near T$_s$. It is possible that the lower dimensionality of thin LCO flakes enables a greater switching control on electrical current, where a larger percentage of the conducting channel is influenced by charge ordering effects than in bulk systems.

It is apparent that being able to study LCO in a thin dimensionality is advantageous. In the initial explorational study, much insight has been gained on some of the fundamental aspects of LCO, among other predictions of a surface 2DEG finding its origin
in topological band structure effects by Radha \textit{et al.}\cite{Radha2021}
as dicussed in Sec. \ref{sec:lco2deg}. Other observations which still bear repeating and understanding, such as superconductivity in swollen LCO, have the potential to benefit from this new approach, to further our understanding of such correlated transition metal oxides.

\section{Conclusion}
In this perspective paper, we have focused on three layered oxide materials, V$_2$O$_5$, MoO$_3$ and LiCoO$_2$, the first two
examples of van der Waals bonded oxides and the latter exhibiting mostly ionic bonding between CoO$_2^{-1}$ layers and Li$^{+1}$
layers. The former can be exfoliated by means of  mechanical exfoliation while the latter requires chemical exfoliation techniques
to obtain atomically thin layers. By carrying out combined studies of their electronic, phonon and transport properties, insights
were gained in how such two-dimensionality affects these oxide's fundamental properties. The effects of phase transitions and ordering
were found to be important in LiCoO$_2$. As is always the case in oxides, point defects and their distribution may play a major role
in the further development and control in such 2D oxides.

From the theory point of view, interesting effects were predicted, for
example phonon frequency shifts, related to changes in 2D screening as well as the breaking of the weak van der Waals bonds.  The electronic
structure of oxides is a complex problem because even small changes may result in strong correlation effects. For instance in
LiCoO$_2$, partial delithiation can break the non-magnetic band filling favored by perfect Li electron donation to the CoO$_2$ layers resulting
in a perfect $d^6$ configuration. Such effects were observed to
occur after annealing of the LiCoO$_2$ nanoflakes resulting in disordering of the remaining Li and Co and formation of more correlated
and lower transport phases. It also leads to interesting magnetic anomalies due to the formation of magnetic moments upon delithiation,
which had been previously observed in bulk but appear to be more prominent in ultrathin 2D LiCoO$_2$. 
In V$_2$O$_5$ and MoO$_3$ which are intrinsically relatively wide gap insulators, n-type conduction can be
caused by oxygen vacancies  or intentional intercalation with alkali or alkaline-earth atoms donating their electrons. This allows
for the fabrication of nanoscale field gate transistor structures and measurements of their intrinsic transport properties. 
However, the intricate band structure of V$_2$O$_5$, which has 1D aspects related to the occurrence of chains within the layers can also
lead to strong anisotropy of the conduction in the plane and intriguing magnetic effects. To fully understand how these differ
in 2D versions, where reduced screening plays a major role, a better understanding of the electronic structure will be required
as well as better control over defects and further refinements of the exfoliation techniques to obtain truly monolayer and large
lateral area materials.

While standard DFT appears to provide reasonable band gaps and band structures, 
the for most semiconductors almost perfectly reliable $GW$ method failed to predict the correct band gaps in V$_2$O$_5$ as well
as in LiCoO$_2$ and the verdict is still out on MoO$_3$. It strongly overestimates the gaps  by several eV. In LiCoO$_2$ surprisingly, one needs
to invoke highly localized Frenkel type excitonic effects to understand the nature of the optical band gap and we expect that the same will be
true in V$_2$O$_5$. Electron phonon coupling effects and in particular polaron formation is also expected to play a major role because
of the large differences in static and high-frequency dielectric constants and the rather flat valence and conduction bands
with high effective masses, which will tend to localize carriers in self-trapped polarons at low temperature. 
Since the self-energy that affects both one-particle spectra and two particle
(\ie optical) spectra is long-range
both the fundamental quasiparticle gap and the optical gap will be strongly affected by 2D induced changes in screening.
While realizing true monolayer 2D physics in these oxides remains challenging, the atomically thin nanostructures studied so far
already hold surprises. The interplay between theory and experiment is important. The computational study of 2D Li-covered CoO$_2$ layers
inspired by the exfoliation experiments revealed the presence of a spin-polarized 2DEG on their surface mediated by the Li bands
coming down in energy and this in turn was shown to be evidence of at least partially covalent bonding between  Li and CoO$_2$ layers
which has topological consequences at the surface. In turn the bonding of Li (or Na) to CoO$_2$ layers in bulk undergoes a symmetry breaking
transition when the layers are pulled apart beyond a critical distance and may lead to essentially a stack of monolayers still associated
with Li or Na on one side and a Li or Na free surface on the other side. The presence of an electron gas formation under these circumstances
may shed important new light on the superconductivity in hydrated (and hence layer expanded) Na$_x$CoO$_2$. Our detailed studies of the
exfoliation process itself as well as the effects of annealing reveals that chemical exfoliation is a complex process with many
still to be fully understood aspects.  Many dots remain to be connected. 
As an outlook, we may expect much new and interesting physics from oxide
systems when they finally break into the world of 2D monolayer materials.  Their rich crystal structures and types of bonding
and spin-dependent and correlated electronic structure will no doubt hold surprises for future research. 

\acknowledgements{This work was supported by the US Air Force Office of Scientific Research
  under grant No. Grant No. FA9550-18-1-0030.  
  The calculations made use of the High Performance Computing Resource in the Core Facility for Advanced Research Computing at
  Case Western Reserve University. We would like to acknowledge Mourad Zeynalov for his experimental contributions throughout the $\alpha$-Na$_{0.96}$V$_2$O$_5$ project. We thank Kenta Kimura and Tsuyoshi Kimura for providing NaV$_2$O$_5$
  crystals.}

{\bf Data vailability:} The data that support the findings of this study are available from the corresponding author upon reasonable request.

\bibliography{lco,lmto,gw,dft,moo3,v2o5,nav2o5,defects,MoO3Transport,kylesrefs}
\end{document}